\documentclass[11pt]{article}

\begin{document}

\title{\huge  The Development of Elementary Quantum Theory from 1900 to 1927 }
\date{}
\maketitle

\begin{center}
{\large Herbert Capellmann}\\
Potsdam, and Inst. Theor. Physik, RWTH Aachen\footnote {capell@physik.rwth-aachen.de}\\
\end{center}

\begin{abstract}

Planck's introduction of the quantum of action in 1900 was followed by 25 years of trial and error in quest of the understanding of the quantum world; different ideas and directions had to be pursued until the path leading to the elementary quantum theory was discovered. Radical changes away from traditional perceptions about natural phenomena were necessary, the entire system of basic concepts in classical physics had to be abandoned and replaced by a new mode of thought. Continuity and determinism of classical laws were no longer applicable on the quantum scale, where dynamical behaviour proceeds by discontinuous and statistical quantum transitions. Albert Einstein laid the essential foundations for the new concept; Max Born made the decisive step further leading to the breakthrough in 1925. The development of the ideas, which eventually resulted in the elementary quantum theory in 1925/26, will be described, relying on original publications and letters written during that period in time by the major contributors. 
The fundamental laws of Quantum Theory derived by Max Born and Pascual Jordan  may mathematically be represented in many different ways, and particular emphasis is given to the distinction between physical content and mathematical representation.\\

\end{abstract}

\newpage

\begin{section}
 {The fundamental differences between classical and quantum physics}
\end{section}

The basic laws of classical physics relied upon the principle "Natura non facit saltus" (nature does not make jumps), transmitted from ancient philosophy. The underlying assumption was the existence of a space-time continuum and all changes in nature should occur continuously within this space-time continuum. Starting towards the end of the 17'th century, the classical laws governing these changes were expressed in form of differential equations or variational principles, where infinitesimally small changes of various physical variables are related to each other. 
Typically these differential equations of classical physics possessed exact solutions for given initial and boundary conditions, at least “in principle”. This led to the general conclusion that nature is “deterministic”;  the state of nature at any given time was believed to be related in a unique way to its state at any past or future time. Even if the development of statistical thermodynamics related probabilities to thermodynamic variables, these probabilities were meant to describe insufficient knowledge of details due to the large numbers of microscopic particles involved, but deterministic behavior of all individual processes was not questioned. \\
{\bf Classical physics} relied upon the principles of {\bf continuity and determinism}: \\
{\bf \noindent
- $\;\;$ Changes of all physical quantities  occur continuously in space and time.
- $\;\;$ The laws determining these changes are deterministic.}\\

When the microscopic "quantum world" was explored towards the end of the 19'th and the beginning of the 20'th century, the basic laws of classical physics (mechanics, electrodynamics and thermodynamics, which had led to great scientific and technological advances during the 19’th century) turned out to be unable to describe the observations. The keys, which eventually should lead to the development of the elementary Quantum Theory by Max Born, Werner Heisenberg and Pascual Jordan in 1925, are contained in:\\
The basic principles of {\bf Quantum physics}:\\
{\bf \noindent
-  $\;\;$ On the microscopic level all elementary changes in nature are discontinuous, consisting of quantized steps: "quantum transitions". \\
-  $\;\;$ The occurrence of these quantum transitions is not deterministic, but governed by probability laws.}\\

Continuity and determinism had to be abandoned, which amounted to a radical change away from all traditional concepts about the laws of nature, and it is not surprising, that it took several decades from Planck's quantum hypothesis in 1900 until Born, Heisenberg, and Jordan formulated the elementary Quantum Theory in 1925.\\
25 years actually are a rather short time, for a totally {\it "new mode of thought in regard to natural phenomena"} (Max Born) to develop, and the new theory of Born, Heisenberg, and Jordan was received with great skepticism by a large fraction of the "physical establishment", e. g. Max Planck, Albert Einstein (although Einstein himself had laid the most important foundations for the new concept) and many others.\\

\begin {section}
{Planck, Einstein, and the "Old Quantum Theory" from 1900 to 1924}
\end {section}

\begin{subsection}
 {Planck's quantum hypothesis }
\end{subsection}

At the meeting of the Deutsche Physikalische Gesellschaft on December 14'th 1900 Max Planck presented the derivation of his radiation law based on a quantum hypothesis and defining a new universal constant, the quantum of action $h$,  now called Planck's constant {\bf (Verh.D.Phys..Ges. 2,553-563,1901)}. Planck's aim had been to explain the spectral distribution of electromagnetic radiation in thermal equilibrium with a surface or a gas of given temperature. 
The quest for the understanding for this "Normal Spectrum" will be decisive for the development of quantum theory. In 1859 Kirchhoff had concluded that the basic laws of thermodynamics required the spectral distribution of radiation in equilibrium to be a universal function of frequency and temperature alone. 
In 1893 Wilhelm Wien formulated his "displacement law" for the radiation energy $u$ as function of 
frequency $\nu$ and temperature $T$:  $u(\nu,T) = a\nu^3 f(\nu/T)$ (Berichte der Berliner Akademie of 9 Feb 1893), and in 1896 experimental studies led Wien to propose a special form, 
"Wien's distribution law": $u(\nu, T) = a\nu^3 e^{-c\nu/T}$ (Ann. d.  Phys. u. Chem. 58, 662, 1896). Already before December 1900  Max Planck tried to derive the special form of the "Normalspektrum" relying only on the phenomenological laws of thermodynamics. First he aimed for the derivation of Wien's special form, the distribution law; when experimental studies showed, that this distribution law gave too low an intensity for low frequencies, Planck extended his phenomenological arguments - by what he himself called an arbitrary assumption - to arrive at (Verh. d. Phys, Ges. 2, 202, 1900)\\

\begin{equation}
u(\nu, T) = a\nu^3 \frac {1}{e^{b\nu/T} - 1}.
\end{equation}

Planck was not satisfied with his "arbitrary assumption"; using the methods of statistical thermodynamics and Boltzmann's relation between entropy and probabilities ($S = k log W$) he obtained the correct functional form of the radiation law:\\

\begin{equation}
u(\nu, T) = \frac{8 \pi h \nu^3}{c^3} \frac {1}{e^{h\nu/kT} - 1} \; ;
\end{equation}

$c$ is the velocity of light, $h$ and $k$ were called "universal constants";  fitting the experimental data available at that time, their numerical values were obtained.\\

Although the discovery of the radiation law and the introduction of the quantum of action remain to be Planck's lasting and immense merit, Planck's physical assumptions for this hypothesis were incorrect.  Planck did not question the purely classical nature of electromagnetic radiation, in Planck's view completely described by Maxwell's equations; electromagnetic radiation and the mechanical behavior of particles (electrons) should still be determined by classical laws. Planck sought the reason for the appearance of the new universal constant in the microscopic interaction process of {\it "elementary resonators"} with the electromagnetic field. Planck suggested that this interaction emitting and absorbing radiation might not be described by classical laws and should be modified, such that the energy of the "elementary resonators" of frequency $\nu$ were restricted to integer multiples of $h \nu$. Energy and entropy exchange between the "resonators" and the electromagnetic field then should determine the spectral distribution of the radiation field.\\

\begin{subsection}
 {Einstein's quanta  of radiation}
\end{subsection}

It was Albert Einstein, who recognized that the electromagnetic field itself is quantized. On 17 March 1905 he submitted the paper entitled {\it "\"{U}ber einen die Erzeugung und Verwandlung des Lichtes betreffenden heuristischen Gesichtspunkt"} ("On a heuristic point of view concerning the creation and conversion of light"), which was published in {\bf Ann. Phys. 17, 1905, 132 - 148}. Einstein (re-)introduced the particle concept of radiation, claiming that - on the microscopic level - light consists of quanta having particle properties.  Einstein concluded that the radiation energy consists of\\
 {\it "...  einer endlichen Zahl von in Raumpunkten lokalisierten Energiequanten, welche sich bewegen, ohne sich zu teilen und nur als Ganze absorbiert und erzeugt werden k\"{o}nnen" (a finite number of energy quanta which are located in points of space, which move without splitting up and which can only be absorbed and created as a whole)}.\\

Einstein based his arguments on the experimental information available about the interaction of electromagnetic radiation with matter, in particular experiments on the creation of cathode rays by light (the "photo-electric effect"), the inverse process of cathode luminescence, the ionization of gases by ultraviolet radiation, and photoluminescence. In all these processes energy is exchanged between radiation and point like {\bf particles}. The particle energies in these processes do not depend on the intensity of radiation but on its frequency $\nu$, whereas only the number of particles involved depends on the intensity. Einstein drew the conclusion that the radiation energy cannot be distributed continuously, but is distributed discontinuously in space in units of $\epsilon = h\nu$, in the same way as matter is made up of discrete particles. Einstein  supported this "heuristic point of view" with a theoretical analysis of the thermodynamic properties of radiation. He established that the "classical limit" of Planck's radiation law for low frequency (i. e. high radiation intensity) as well as the limit of high frequency (i. e. low radiation intensity, where Wien's distribution law is applicable) confirmed the existence of "light quanta", the thermodynamic properties of radiation being consistent with the behaviour of an ideal gas of particles. \\

Einstein was well aware that, introducing the particle concept of light quanta, he not only solved the problem of the interaction process of light with matter, he also created a new problem: If light consists of particles, how can "wave optics" be reconciled with the particle picture? Einstein himself pointed out that his particle concept of light left the question open why "wave optics" gave an accurate description of many macroscopic phenomena. Einstein stated explicitly, that \\
{\it "...die Undulationtheorie des Lichtes sich zur Darstellung der rein optischen Ph\"{a}nomene vortrefflich bew\"{a}hrt hat und wohl nie durch eine andere Theorie ersetzt werden wird" (to describe purely optical phenomena the undulation theory of light has proven its worth excellently  and most probably will never be replaced by another theory)}. \\
Remarkably, Einstein points already into the correct direction to solve this new problem, suggesting that wave optics should apply to averages only. During the years before 1905, Einstein had made fundamental contributions to a similar problem of thermodynamics (A. Einstein, Ann. d. Phys. 9, 417-433, 1902; 11, 170-187, 1903; 14, 354-362, 1904): How to connect the individual properties of the constituents of matter to the macroscopic laws of phenomenological thermodynamics. Atoms in a gas have energy and momentum; temperature, entropy, pressure are properties of macroscopic averages. Gibbs and Einstein established the connection, based on Boltzmann's relation between probabilities and entropy. Einstein recognized that the equivalent problem had to be solved to reconcile particle properties and wave phenomena of radiation. But the solution to "Einstein's new problem", the connection between the microscopic particle behaviour of light and the phenomenology of wave optics, will have to wait until the advent of a valid quantum theory several decades later.\\

Planck, as well as the vast majority of his contemporaries up to and partly beyond 1925, rejected Einstein's revolutionary concept.
From today's perspective it might seem highly peculiar, that Einstein's hypothesis was met with general disbelief for several decades, but we should not underestimate the highly revolutionary character of Einstein's proposal, which, for Max Planck and most of his contemporaries, seemingly negated one of the most important scientific advances of the 19'th century, Maxwell's theory of electromagnetism and the explanation of all wave-optical phenomena, such as diffraction, dispersion, refraction, and reflection.\\

Let us examine the apparent contradictions between Einstein's hypothesis and the traditional interpretation of radiation in more detail, in particular the central problem of the “spectral distribution” of radiation in thermal equilibrium with matter. The experimental determination of the spectrum directly assigned wavelengths and frequencies - i. e. wave characteristics - to radiation; these experiments seemingly directly disproved Einstein’s hypothesis. Einstein claimed that radiation in thermal equilibrium consists of statistically emitted and absorbed particles, that these particles are noninteracting and having thermal properties of an ideal gas! It seemed inconceivable that noninteracting particles, statistically emitted and absorbed, could provide frequencies and wavelengths; even the basic concept of constructive and destructive interference seemed to be in total contradiction with Einstein's picture.  \\

Einstein conceded, that he did no longer understand the origin of wave optics; but he also concluded that the experimental information about the interaction between radiation and matter was incompatible with wave behaviour on microscopic scales. This belief was strengthened in the following years, when Einstein made further decisive contributions for the development of quantum theory.\\

\begin{subsection}
 {Einstein's further essential contributions}
\end{subsection}

{\bf In 1906 (Ann. Phys. 22, 1906, 180-190; Ann. Phys. 22, 1907, 800)} Einstein applied the quantization concept to oscillations of atoms in solids, necessary for the understanding of thermal properties, in particular the temperature dependence of the specific heat.\\

{\bf In 1909 (Phys. ZS. 10, 185-193, 1909; and Phys. ZS. 10, 817-826, 1909)} he discussed the energy and momentum fluctuations of the electromagnetic field and showed that the experimentally established radiation law was fully compatible with the particle concept of light quanta, carrying energy $\epsilon=h\nu$ and momentum  $\epsilon$/c. It is in this publication that Einstein mentions the momentum of the light quantum explicitly (which he had not done in 1905, although - for Einstein in particular - it was certainly obvious that  pointlike energy quanta moving with the speed of light must possess momenta compatible with relativity theory). Einstein derived the energy and momentum fluctuations of radiation in equilibrium and connected the fluctuation characteristics to the functional form of the radiation spectrum. He showed that the fluctuation formula for energy and momentum based on the particle concept of radiation were fully consistent with the experimentally established form of the radiation law.  He also showed that wave optics, on the other hand, could only describe the energy and momentum fluctuations for the high intensity and low frequency part of the spectral distribution (the classical limit). The experimentally established form of the radiation law for high frequency, however, was shown to be incompatible with wave optics.\\

The second publication mentioned above (Phys. ZS. 10, 817-826, 1909) is a reproduction of a lecture Einstein had given at the annual meeting of the Association of German Scientists (81. Versammlung deutscher Naturforscher und \"{A}rzte, Salzburg, 21 Sept 1909) and concludes with a discussion, where Planck expresses his objections to Einstein's light quanta. Planck confirms that - in his opinion -  Maxwell's equations remain to be  a complete description of electromagnetic radiation; quantum effects should be restricted to the interaction of matter and radiation. Emission and absorption processes should require strong acceleration of "resonators"  which should fall outside the traditional description and that these poorly understood processes should be responsible for restricting the resonator energies to integer multiples of elementary quanta. \\

{\bf In 1916/17 (Verh.d.D.Phys.Ges. 18, 318-323, 1916; and Phys. Z. 18, 121-128, 1917)} Einstein established the conditions necessary for thermal equilibrium between radiation and matter. It is this publication, which will be of decisive importance for the Quantum Theory to be developed by Max Born, Werner Heisenberg , and Pascual Jordan. Einstein gives the first derivation of the complete form of Planck's radiation law, which was based on physical arguments fully consistent with the future development of Quantum Theory. Furthermore he confirmed the existence of the photon momentum and its necessity for the establishment of thermal equilibrium between radiation and a gas; energy and momentum conservation for all individual processes are required to assure that thermal equilibrium is maintained. \\

Einstein introduced the concepts of spontaneous emission and of stimulated absorption and emission. Due to energy conservation in all individual processes, spontaneous creation of a photon is possible only if accompanied by a quantum transition from a higher to a lower energy state; stimulated processes are possible for absorption and emission, their {\bf probability} being proportional to the intensity of the applied field. Einstein introduces {\bf transition probabilities}, a term, which will play a crucial role in the development of Quantum Theory in 1925. But it is important to notice the distinction between Einstein's use of "probabilities" and the meaning assigned to {\bf transition probabilities} in the future Quantum Theory. In Einstein's view, "probabilities" reflect thermodynamic arguments. If we want to connect phenomenological laws of thermodynamics with the underlying  microscopic processes, it is typically sufficient to restrict the microscopic description to "probabilities", the phenomenological laws on macroscopic scales being independent of a detailed knowledge of all microscopic processes involved. Einstein maintained that the microscopic behaviour should be governed by deterministic laws in principle; concerning his own theory he conceded that {\it "the theory has the following weaknesses: First, it does not provide a connection to wave theory; second, it leaves time and direction of the elementary processes to chance"}.

Later,  Max Born will make the decisive step further; he will conclude that discontinuous and statistical behaviour is a fundamental property of nature for all elementary processes, classical laws being valid approximately for macroscopic averages only. This principle will be essential for the Quantum Theory of Born, Heisenberg, and Jordan developed in 1925, which will assign purely statistical laws to all individual quantum transition.  Einstein, however, will refuse to take this step. \\

Einstein's conviction concerning the purely particle character of radiation quanta even led him to predictions about totally new quantum phenomena, which were to be confirmed only much later. In 1924 N. Bose sent a manuscript to Einstein, which opened the new field of quantum statistics, in particular Bose statistics. Einstein recognized the importance of Bose's discovery, he translated the paper himself\footnote {the paper finishes with: "translated by A. Einstein."} and had it published {\bf (S. N. Bose, "Plancks Gesetz und Lichtquantenhypothese", Z. Phys. 26, 178-181, 1924)}.  Bose's publication contained the first direct derivation of Planck's radiation law, which connected exclusively particle properties of photons directly to the thermodynamic properties of 
radiation\footnote{Einstein's own derivation of 1916 still had to use experimental information from Wien's displacement law}, at the expense of requiring new quantum statistical principles:\\
a) Phase space (e.g. spanned by position and momentum variables) is divided into unit volumes $h^3$; and \\
b) physical states are characterized by the occupation numbers of the various unit volumes $h^3$.\\
The first principle might be called "phase-space quantization"; a particle is no longer characterized by {\bf precise} values of position and momentum, the uncertainty within the finite volume $h^3$ represents a {\bf quantum uncertainty}\footnote{"Fermi statistics" (E. Fermi, Z. Phys. 36, 902, 1926, rec 24 Mar 1926) uses the same argument.}. 
The second principle implies {\bf indistinguishability} of identical particles. These principles will turn out to be of general validity; "Bose-statistics" allows multiple occupation numbers. \\ 

Two publications by Einstein followed quickly {\bf (Sitz. Berlin Ak. d. Wiss. 10-07-1924, and  08-01-1925)}, which applied Bose statistics to the Quantum theory of ideal gases. Einstein stressed that Bose's concept implied the {\bf indistinguishability} of identical particles and necessarily leads to new consequences and correlation effects even without interaction.
The particle character of photons should find its complete analogy in the behaviour of other particles, and Einstein applied the highly peculiar consequences of  Bose statistics to particles with finite mass, predicting the possibility of the macroscopic quantum phenomenon of (Bose-) Einstein Condensation ("BEC"). It should take 70 more years until BEC could be realized experimentally. Einstein extended the analogy between light quanta and other particles further, suggesting, that "interference effects" similar to "wave optics" should be expected for  particles with finite mass as well\footnote {In this context Einstein refers to the thesis of Louis de Broglie, of which Einstein had obtained a copy prior to publication. De Broglie's concept of "phase-waves" will be discussed later in the chapter about "Wave Mechanics".}.\\

\begin{subsection}
 {The "Old Quantum Theory"}
\end{subsection}

Planck's concept, i. e. classical radiation and continuous electronic oscillations emitting and absorbing classical radiation in some mysterious and unexplained way, formed the basis for the "old quantum theory", which dominated quantum theoretical attempts from 1913 up to the year 1925. 
The attention shifted away from Planck's radiation law to the analysis of spectroscopic data. In 1908 Walter Ritz (Phys. Zeitschr. 9, 521, 1908) discovered that the multitude of observed spectral frequencies  characteristic of atoms and molecules could be classified by a combinational principle:
 $\nu_{ij} = X_i - X_j$. The Balmer-Rydberg series of hydrogen
($\nu_{nm} = X(1/n^2 - 1/m^2$), n and m integers) 
was a special example.  Although Einstein's light quanta allowed a simple explanation, this was rejected; radiation was considered to be classical. \\

In 1913 Niels Bohr proposed an atomic model {\bf (Phil. Mag. 16, 1-25, 1913)}, which was based on a physical picture similar to the planetary system, electrons taking the place of planets and the nucleus replacing the sun. Electronic orbits were determined  by classical equations of motion.
The free radiation field was assumed to be classical, described by Maxwell equations. Bohr rejected Einstein's light quanta and the quantum nature of the electromagnetic field; repeatedly Bohr insists on the {\bf continuous} character of the field, which - in his mind - was the only possibility to explain wave optical phenomena. Like Planck, Bohr considered the interaction process between electrons and radiation to be responsible for quantum effects, not compatible with classical physics.\\
 
Quantum conditions were postulated for the selection of stationary states, i.e. stable atomic configurations, and for the emitted radiation frequencies. Within stationary states electrons were assumed to perform  orbits of circular or elliptical shape with rotational frequencies $\omega_n$. No radiation was to be emitted within these stationary states; emission of homogeneous radiation should result from transition processes between two stationary states. \\

 Multiplication of Ritz' combinational  scheme   by Planck's constant resulted in Bohr's quantum condition for the emitted radiation frequencies\\

\begin{equation} 
h\nu_{ij} = W_i - W_j.
\end{equation}
\noindent
Bohr identified the right hand side as the difference in energy of two stationary states. \\

We stress the difference in interpretation of this equation,
which is used by both Einstein and Bohr, but has different physical significance according to  Einstein and  Bohr. They agree on the right hand side, which is the energy difference between two states, but they have totally different interpretations for the left hand side.  
For Einstein $h\nu_{ij}$ is the energy of the photon, i.e. a particle, emitted or absorbed by the transition process; the equation expresses energy conservation. As mentioned before, Bohr was convinced that radiation is classical; only a wave picture should be able to account for the observation of interference effects.
Since Bohr's calculations will produce rotational frequencies $\omega_n$  without any  relation to the observed radiation frequencies, the condition $h\nu_{ij} = W_i - W_j$ constituted an additional assumption with no apparent physical explanation.  Bohr "solved" this problem suggesting "virtual oscillators" responsible for emission and absorption of the observed frequencies. \\

The first aim was to give an explanation for the Balmer-Rydberg series of hydrogen ($\nu_{nm} =  X(1/n^2 - 1/m^2)$, n and m integers); the frequency condition (eq. 3) required the $W_n$ to be given by $W_n = h X/n^2$.
To obtain the desired result, Bohr made the following assumption: \\
i) The electron is circling the nucleus with rotational frequency $\omega_n$; classical mechanics relates $\omega_n$ to the energy $W_n$. \\
ii) The selection of "stationary states" requires the $W_n$ to be equal to $n$ times the energy quanta of  
"Planck oscillators"\footnote{The "Planck oscillators"  are different from the "virtual oscillators"; the latter are supposed to have frequencies equal to radiation frequencies $\nu_{nm}$. The $f_n$ are unrelated to the $\nu_{nm}$.}
 of  frequencies $f_n$, i.e. $W_n = n h f_n$. \\ 
iii) To complete the determination of the $\omega_n$ and $W_n$, a relation between oscillator frequencies $f_n$  and rotational frequencies $\omega_n$ was required. The relation $f_n = \omega_n/2$ resulted in

\begin{equation}
W_n = \frac{2 \pi^2 m e^4}{h^2} \frac{1}{n^2}.
\end{equation}
\noindent
Inserting these values into the frequency condition ($h\nu_{nm} = W_n - W_m$) reproduced the experimentally observed frequencies of hydrogen. Classical equations of motion were also used to calculate radii and angular momenta of stationary states. The results for electronic radii agreed with experimental estimates of atomic radii. The angular momentum of state $n$ was obtained as $J_n = n \frac{ h}{2\pi}$. \\

Bohr's calculations were partly "successful" due to dimensional reasons. If an energy scale is to be obtained from the available natural constants (electron mass, elementary charge, and Planck's constant), the combination $m e^4 / h^2$ follows. Integer numbers were obtained invoking energy quanta of Planck oscillators; the factor $1/2$ between $f_n$ and $\omega_n$  was a free parameter and could be chosen to obtain the correct energies, although all physical assumptions were incorrect. \\

While the heuristic assumptions at first sight  seemed to be rather arbitrary, the "success" to reproduce the observed spectral lines for atomic hydrogen was taken as a first step towards  the solution of the quantum mystery; similar attempts followed to reproduce the observed spectral lines of other elements. Additional assumptions concerning the nature of multielectron orbits were added; electrons were positioned in symmetric arrangements (all of them classically unstable, however) on one or more rings or ellipses rotating rigidly. \\

The theoretical activities in quantum theory for the next 12 years were dominated by extensions of  Bohr's concept.  Sommerfeld (Ann. Phys. 51, 1-94, 1916) proposed a reformulation of the quantum condition: {\it The integral of position times momentum over one period is required to be equal to an integer multiple of $h$}, which led to further application, e.g. vibrations in molecules and solids. The majority of contributions, however, were directed towards a more detailed analysis of spectroscopic data. \\ 

The review by Bohr {\bf (Z.Phys.13,117-165, 1923)} restates the basic assumptions of the "old quantum theory":\\

\noindent
- Classical continuity in space and time is retained for all processes.\\
- The electromagnetic field is purely classical.\\
- Periodic orbits of electrons around nuclei are determined by classical equations of motion.\\
- Stationary states are selected by quantum conditions: The integral of position times momentum over one period is required to be equal to an integer multiple of $h$.\\
- No radiation is emitted within stationary states.\\
- "Virtual oscillators"  - with frequencies unrelated to the periodicity of the orbits - emit or absorb  homogeneous radiation during transitions between stationary states.\\
- The frequencies of the "virtual oscillators" and the radiation frequencies are equal to the energy differences between stationary states divided by $h$.\\
- The transition process itself is continuous having finite duration.\\

The discovery of the Compton Effect (A. H. Compton, Bull. Nat. Res. Council 4, Nr 20, Oct 1922) and its theoretical interpretation based on Einstein's photon concept (H. A. Compton, Phys. Rev. 21, 483, 1923, and P. Debye, Phys. Z. 24, 161, 1923) generated some support for Einstein (e. g. W. Duane\footnote {William Duane was the first to point out that Einstein's particle concept of photons is able to describe the so called "wave optical" observations, which had been interpreted to result from "interference phenomena". Duane's paper will be described later in more detail} , Proc.Nat.Ac.Sc, 9, 158, 1923; W. Pauli, Z. Phys. 18, 272, 1923, and Z. Phys. 22, 261, 1924), but the majority - and in particular Niels Bohr and his followers - remained unconvinced.\\

The supposedly purely classical nature of the electromagnetic field is also the stumbling block for
N. Bohr, H. Kramers, J. C. Slater (the "BKS theory"; Phil. Mag. 47, 785-802, 1924; and Z. Phys. 24, 69-87, 1924), when they try to  address the interaction process between radiation and matter in more detail. To justify their assumptions about the continuous and classical character of  radiation on one side and the fast transition process between stationary states on the other, additional assumptions are introduced, which - from today's perspective -  seem rather bizarre. As before, two different time scales are used, continuity in time is retained for both scales: a "classical" time scale for classical radiation and a "fast" time scale for the finite duration of the transition process between stationary states, associated with "virtual oscillators".  Emission of continuous radiation at the oscillation frequency is postulated to be a collective phenomenon involving many atoms simultaneously. An additional radiation field is introduced ("virtual radiation") instantly connecting different atoms and coordinating transitions in distant atoms such that the continuous and classical behaviour of electromagnetic radiation is maintained.  Energy and momentum conservation are claimed to be absent for individual transitions and are postulated to be valid on average only. \\

The "BKS theory" marks the impasse in which the "old quantum theory" had ended up. The hypothetical electron dynamics in atoms was without any experimental evidence; radiation was still assumed to be classical; classical equations of motion were supplemented by heuristic and unrelated hypotheses; a path towards a general and internally consistent quantum theory seemed not in sight. A fundamentally new concept  was necessary to attain the understanding of the quantum world.\\

\begin{section}
 {Born's discontinuous "Quantenmechanik"}
\end{section}

 The basic ideas about discontinuous and statistical behavior of all elementary processes in nature are due to Max Born.  Already several years before 1925 Born questioned the applicability of the classical concepts of continuity and determinism, as is evident in his correspondence with Pauli and Einstein. A letter to Pauli of 23 december 1919 contains \\  
{\it "man darf die Begriffe des Raumes und der Zeit als ein 4-dimensionales Kontinuum nicht  von der makroskopischen Erfahrungswelt auf die atomistische Welt \"{u}bertragen, diese verlangt offenbar eine andere Art von Mannigfaltigkeit als ad\"{a}quates Bild"}\\
({\it One should not transfer the concept of space-time as a four-dimensional continuum from the macroscopic world of common experience to the atomistic world; manifestly the latter requires a different type of manifold.})\\
Born's correspondence with Einstein from early 1920 onwards reflects Born's conviction (and Einstein's 
hesitance\footnote{Einstein's letter to Born of 27 Jan 1920: {\it "Daran, dass man die Quanten l\"{o}sen m\"{u}sse durch Aufgeben des Kontinuums, glaube ich nicht"} (I do not believe that the continuum has to be abandoned to solve the quantum problem.)}) 
to accept discontinuous and statistical behavior on atomic scales. But it should take several years until this conviction led to a valid theory.  The concept of discontinuous and statistical {\it "quantum transitions" ("Quantenspr\"{u}nge")} grew slowly during several years until Born, Heisenberg, and Jordan developed the elementary quantum theory during the year 1925. \\

The publication  "Quantentheorie und St\"{o}rungsrechnung" (M. Born, Naturwissenschaften 11, 537-542, 1923) confirmed Born's conviction that a radical change was required away from classical concepts prevalent in the "old quantum theory". Born concludes:\\
{\it "Es wird immer wahrscheinlicher, dass nicht nur neue Annahmen im gew\"{o}hnlichen Sinne physikalischer Hypothesen erforderlich sein werden, sondern dass das ganze System der Begriffe der Physik von Grund auf umgebaut werden muss."}\\
({\it "It becomes more and more probable, that not only new assumptions in the usual sense of physical hypotheses are needed, but that the entire system of basic concepts in physics will have to be rebuilt radically."})\\

During the winter semester 1923/24 Born held a series of lectures on "Atommechanik", with the aim to explore the limits of the "old quantum theory" and to define a program to go beyond these limits towards the {\it "endg\"{u}ltige Atommechanik"} ("final atom mechanics"). Werner Heisenberg had just arrived in G\"{o}ttingen to take up the post of Born's  "Assistent", Pascual  Jordan was still completing his doctoral thesis (the doctoral exam taking place in the spring of 1924).  Born's lecture notes were published in the book "Vorlesungen \"{u}ber Atommechanik, 1. Band" ("Lectures on Atom Mechanics, 1st Volume"); Springer Verlag, November 1924. The future 2nd volume, should (so the very optimistic announcement) contain the "final atom mechanics".\\

 Born accepts  the validity of classical mechanics and electrodynamics for {\bf macroscopic processes} only, and recognizes their failure for the understanding of the quantum world. He criticises that the "old quantum theory" conserves concepts of classical mechanics, in particular the continuous movements of electrons, which are not accessible to observations ({\it "Es scheint, dass diese Gr\"{o}ssen prinzipiell der Beobachtung nicht zug\"{a}nglich sind".}) Born concludes that the "old quantum theory" provides a formal calculational scheme, applicable to special cases only, but does not contain the "true quantum laws". He specfies that {\bf the "true quantum laws" should contain relations between quantities, which are observable} ({\it "Von diesen wahren Gesetzen m\"{u}ssen wir verlangen, dass sie nur Beziehungen zwischen beobachtbaren Gr\"{o}ssen enthalten"}).\\ 

 The path towards the future Quantum Theory is defined as \\

{\it  "the systematic transformation of classical mechanics into a discontinuous atomic mechanics.....} \\

{\it the new mechanics replaces the continuous manifold of} (classical) {\it states by a discrete manifold, which is described by "quantum numbers"....\\

quantum transitions between different states are determined by probabilities...\\

the theoretical determination of these probabilities is one of the profound tasks of  Quantum Theory...."}.\\

Continuous electronic orbits, which had been at the center of Bohr-Sommerfeld theory, were not observable, experiments showed sharp spectral lines, which could be interpreted resulting from electronic transitions. These transitions appeared to be sudden and occurring statistically, and Born was willing to accept that discontinuous and statistical behavior is indeed governing the laws of nature on microscopic scales. Born's correct interpretation of the available experimental data  preceded the successful development of the theory; and it is quite logical that the mathematical formulation developed by Born, Heisenberg and Jordan of the new fundamental laws had a discontinuous form, "matrix mechanics".\\

 The first concrete step to abandon the space-time continuum and replace it by a "new type of manifold"  is contained in\\
{\bf   "\"{U}ber Quantenmechanik" (Max Born, Z. Phys. 26, 379-395, 1924)}.\\
Born's line of attack on the quantum problem is essentially different from the "old quantum theory":  Whereas the Bohr-Sommerfeld approach used Planck's constant to obtain stationary states, leaving the transition processes still open, Born directly attacked the dynamic behavior, i. e. the {\bf quantum dynamics} describing transitions between different states.
 It is in this publication that Born introduces the term "Quantenmechanik", which indicates that the mechanical behaviour is not continuous on the elementary level; all changes  proceed in quantized, i.e. discontinuous, steps. Born starts from quantization of action, which he accepts as established by experiment. Born's central argument will be, that quantization of action in terms of Planck's constant $h$ will change the continuous variation of classical variables  into {\bf discontinuous quantum transitions}, requiring all physical variables to change discontinuously. \\

To put this idea into mathematical form, Born starts from classical mechanics using the Hamilton-Jacobi formalism, where "action" can be used as physical variable, its canonically conjugated variable being a dimensionless "angle-variable". Detailed knowledge of the Hamilton-Jacobi formalism was not commonplace at the time (nor today), and its use made Born's publications difficult to read. Born had obtained his Doctorate and Habilitation in mathematics and he used the mathematical technique most useful for the physical problem to be solved; "action" as physical variable was particularly suited, because the central idea,{\bf "changes of action variables are restricted to integer multiples of $h$"}, could be implemented directly.\\

The second mathematical technique, essential for the implementation of  Born's concept, is perturbation theory. Classical laws describing dynamics typically are formulated in form of differential equations, where infinitesimally small variations of physical variables are related to each other. Born wanted to replace infinitesimally small classical intervals by their discontinuous quantum equivalents. To achieve his goal, he started from classical lowest order perturbation theory, applicable to infinitesimally small changes; the transition from continuous classical dynamics to  discontinuous "quantum dynamics" is achieved requiring action intervals to be multiples of the finite quantum of action $h$. Infinitesimal changes of other physical variables in classical physics are expressed via their dependence on the action variables and differential quotients are replaced by difference quotients. Classical laws in terms of differential equations are thereby replaced by quantum mechanical difference equations.\\

\noindent
Let us summarize the central idea of "Quantenmechanik" by \\
{\bf "Born's quantization condition":}\\
 All elementary changes occurring in nature must be discontinuous, because the action variables may change by integer multiples of Planck's constant only; the discrete behavior of  action variables will affect all other variables as well.\\

\begin{section}
 {The Quantum Theory of Born, Heisenberg, and Jordan}
\end{section}

Although Born's first publication "\"{U}ber Quantemechanik" did not carry the discretization of nature very far, a complete elementary Quantum Theory followed during the following year, developed in a series of  four publications discussed below.\\

\begin{subsection}
 {M. Born and P. Jordan, Z. Phys.33, 479-505, 1925}
\end{subsection}

 The first important step is contained in the publication by {\bf Max Born and Pascual Jordan (Z. Phys.33, 479-505, 1925; rec. 11-06-1925)}, (referred to as {\bf BJ-1} in the following).\\
The authors state their intention to derive basic quantum laws containing observable quantities only, continuous movements of electrons are to be eliminated. 
Born and Jordan treat the basic problem of the interaction of radiation with matter, in particular the {\bf transition dynamics} of emission and absorption. The starting point is  Einstein's publication of 1916/17 ({\bf A. Einstein, Verh.d.D.Phys.Ges. 18, 318-323, 1916; and Phys. Z. 18, 121-128, 1917}), which they recognize to contain {\it "the basic law of quantum optics" ("Grundgesetz der Quantenoptik")}. 
Born and Jordan  develop {\bf quantum mechanical perturbation theory for the transition processes}, spontaneous emission and induced emission and absorption. \\

They start by {\bf classical} perturbation theory, calculating the effects on electronic properties induced by a time dependent field. The coupling between electromagnetic field and electrons in atoms is restricted to the electric dipole moment (dipole approximation, higher multipole couplings are neglected). 
To make the transition to {\bf quantum mechanical perturbation theory}, Born's discrete kinematics and dynamics is imposed "by hand": The classical mechanical variables are replaced by a discrete manifold of complex {\bf "quantum vectors" } (which we shall later call {\bf "matrix elements"}), discontinuously connecting two quantum states, whose action variables $J$ differ by a finite multiple $\tau$ of  Planck's quantum of action.\\

To make the connection to the following publications easier, specifications concerning the notation used by Born and Jordan are helpful. For the problem of radiative transitions treated  in this publication BJ-1 and the dipole approximation used, the relevant atomic degree of freedom is the electric dipole moment ${\bf p}$. Born and Jordan use a Fourier expansion of ${\bf p}$ and the Fourier coefficients of ${\bf p}$ are denoted by ${\bf A}$\footnote {Born and Jordan use gothic letters to indicate vector quantities, as was usual in Germany at the time}. The "quantum vectors" defined by Born and Jordan are the Fourier coefficients ${\bf A}$, which, according to Einstein's {\it "basic law of quantum optics"}, fulfill the relation\\
\begin{equation}
{\bf A}^{+}_\tau (J) = {\bf A}^{-}_\tau (J + \tau h).
\end{equation}
$J$ is the action variable, $h$ is Planck's constant and $\tau$ is a positive integer. ${\bf A}^{+}_\tau (J)$  is a quantum vector representing a transition increasing the action variable from $J$ to $J + \tau h$, and  ${\bf A}^{-}_\tau (J + \tau h)$ is a quantum vector representing a transition decreasing the action variable from $J + \tau h$ to $J$.\\
The continuous changes of classical action variables are discretized by averaging classical action variables over finite intervals of $h$; in addition to discretization the averaging process introduces a statistical element into the theory. As already in Born's previous publication,  infinitesimal changes of other physical variables in classical physics are expressed via their dependence on the action variables and differential quotients are replaced by difference quotients. The averaging process over finite action intervals of $h$ implies that physical variables do no longer have precise values, only averages over the corresponding action intervals may be defined.\\

Since physical variables may change discontinuously only, time as continuous variable is eliminated as well; time will only be contained implicitly in "probabilities per unit time".  The quantum theoretical probabilities for spontaneous and induced transitions are derived. The guiding principles are that the laws of classical optics  must be recovered for averages; this leads to the requirement:\\
{\bf "Transition probabilities" are proportional to the absolute square of the "quantum vectors"}. \\
Energy conservation for individual transitions is respected, consistent with Einstein's photon picture, the photon carrying an energy $h\nu$; the results for absorbed and emitted radiation energies  calculated classically are recovered for the averages.  \\

Some remarks concerning the concept of "transition probabilities", which Einstein had introduced in 1916/17
(Verh.d.D.Phys.Ges. 18, 318-323, 1916; and Phys. Z. 18, 121-128, 1917):
As pointed out in the discussion of Einstein's publications in chapter {\bf 2.3}, his "probabilities" were elements of the statistical thermodynamic approach to thermal equilibrium between radiation and matter, but  Einstein maintained that the microscopic behaviour should be governed by deterministic laws in principle. Born and Jordan, however, postulate that discontinuous and statistical behaviour is a fundamental property of nature for all elementary processes.\\

\begin{subsection}
 {W. Heisenberg, Z. Phys.33,879-893,1925 }
\end{subsection}

{\bf Werner Heisenberg (Z. Phys.33,879-893,1925; rec. 29-07-1925)} makes the next important step towards the discontinuous quantum theory, following the direction defined by Born-Jordan, but partly retaining Planck-Bohr-Sommerfeld type ideas and methods. Heisenberg had spent the time from late September 1924 in Copenhagen with Bohr and was strongly influenced by Bohr's thinking. The development of Heisenberg's ideas during the critical period between 1924 and 1927 may be traced in his frequent correspondence with Pauli\footnote { Both had been students of Sommerfeld to obtain their doctorates in Munich, and both went to G\"{o}ttingen afterwards to work with Born.} (W. Pauli, Scientific Correspondence, Springer Verlag). Heisenberg returned to G\"{o}ttingen for the start of the summer semester 1925 (beginning of May) and witnessed the development of the paper by Born and Jordan (BJ-1).\\

 Heisenberg's important new idea: To use the quantum vectors (= amplitudes) of Born and Jordan directly for calculations and to use {\bf matrix multiplication rules} for these amplitudes. Heisenberg addresses radiative transitions and following Born and Jordan, Heisenberg repeats the intention to obtain quantum laws containing observable quantities only. The second important part  adopted from Born and Jordan is the partly discontinuous representation of mechanical degrees of freedom. But Heisenberg also retains  essential elements of Bohr's concept.  In Bohr's view radiative transitions between stationary states  were supposed to be triggered by something vibrating inside the atom (Bohr's "virtual oscillators"), emitting and absorbing {\bf continuous} radiation at the oscillation frequencies. Bohr's oscillation concept, continuous in time, is an essential part of Heisenberg's physical picture and interpretation. Heisenberg's mathematical representation of the oscillatory degrees of freedom is a mixture taken from Born-Jordan concerning discontinuous spatial degrees and Bohr's ideas, retaining continuity in time. Since Heisenberg, like Bohr, still assumed that "virtual oscillators" were emitting and absorbing {\bf continuous} radiation, the oscillators had to conserve oscillatory behaviour continuous in time as well. \\

The paper starts with general remarks about the coupling of radiation to electronic degrees of freedom via dipole and higher multipoles, which leads  to the question, how to represent products of classical variables by appropriate quantum theoretical quantities. Heisenberg proposes  time dependent amplitudes  ${\bf A}( n, n - \alpha ) e^{i\omega (n, n-\alpha)t}$, where the $\omega (n, n-\alpha)$ are given by $\frac{1}{\hbar} \;(W(n) - W(n - \alpha))$. The $W(n)$ will become the energies of  stationary states.  Heisenberg's complex amplitudes vectors ${\bf A}( n, n - \alpha)$ correspond to the quantum vectors of Born and Jordan representing the Fourier amplitudes of  the electric dipole moment. ($n$ and  $ n - \alpha$ characterize different quantum states, Heisenberg's $\alpha$ corresponds to $\tau$ in BJ-1). Heisenberg's technical novelty consists in the use of matrix multiplication rules for the amplitudes; due to the special form of the frequencies $\omega (n, n-\alpha)$, matrix multiplication rules apply to the amplitudes directly, irrespective of the time dependent factor. \\

The actual calculations carried out are for a single oscillator degree of freedom. Heisenberg represents the position variable by spatially discrete amplitudes depending on time in continuous fashion:
$a( n, n - \alpha ) e^{i\omega (n, n-\alpha)t}$. These amplitudes are inserted into the classical oscillator equation 
$\ddot x + \omega_0^2 x + \lambda x^3 = 0$; the nonlinear term will be treated in perturbation theory. Stationary states are selected by a modified Bohr-Sommerfeld quantization recipe, relying on continuity in time: The integral of $m\dot x^2$ over a period in time is required to be equal to Planck's constant.  The modification is due to W. Thomas (Naturwissenschaften 13, 627, 1925) and W. Kuhn (Z.Phys. 33, 408, 1925) in the attempt to describe dispersion; two separate oscillators are used for transitions upwards and downwards in energy, contributing with opposite sign to the quantization integral. The solution determines energies and amplitudes. Concerning the energies obtained, the Thomas-Kuhn modification of the Bohr-Sommerfeld quantization recipe leads to the correct quantum mechanical eigenvalues of the linear oscillator, including the zero point energy.  \\

Heisenberg's method  retains essential defects of the "old quantum theory":  The  Bohr-Sommerfeld quantization determines stationary states corresponding to classically periodic states only, a general quantization scheme is not available. The determination of  stationary states corresponds to static properties, the dynamics of transitions between different stationary states is not addressed. More generally, "quantum theoretical equations of motion", i. e. the equivalent of the classical equations of motion to describe "quantum dynamics" of transitions, are absent.  Relying on Bohr's concept of "hidden oscillators" emitting continuous radiation, Heisenberg makes the erroneous assumption that the "virtual oscillator" dynamics generates radiation at the corresponding frequency. Although Heisenberg does not treat the coupling between electronic degrees of freedom and the electromagnetic field explicitly, he nevertheless argues that the squares of the oscillator amplitudes directly determine the probabilities for radiative transitions. \\

 Heisenberg himself was well aware of the limitations, as his letter to Pauli of July 9, 1925 (just before his paper was submitted) testifies: He is convinced that the critical part (expressing the opinion that electronic movements are not observable, whereas proper quantum laws should contain observable quantities only), is fully justified, but that his "positive part" (his explicit treatment of quantum oscillations) is rather formal and scanty, but could serve others to proceed further: ({\it "...dass ich aber den positiveen f\"{u}r reichlich formal und d\"{u}rftig halte; aber vielleicht k\"{o}nnen Leute, die mehr k\"{o}nnen, etwas Vern\"{u}nftiges draus machen"}). \\

\begin{subsection}
 {M. Born and P. Jordan, Z. Phys. 34, 858, 1925 }
\end{subsection}

Heisenberg's paper does indeed contain essential indications how to proceed further, which will be used by Born and Jordan to achieve {\bf the final breakthrough}:\\
{\bf M. Born and P. Jordan (Zur Quantenmechanik; Z. Phys. 34, 858-888, rec. 27-09-1925)}, (referred to as {\bf BJ-2} in the following) obtain {\bf the new fundamental laws: commutation relations and quantum theoretical "equations of motion"}.\\
 
Let us compare the methods and results of Born and Jordan in the preceding publication BJ-1 to those of Heisenberg. In BJ-1 Born and Jordan {\bf do} address the "quantum dynamics" of discontinuous quantum transitions, and they {\bf do} have a generally valid quantum condition, "Born's quantization condition": Changes of action variables are restricted to integer multiples of the quantum of action $h$, implying that all physical variables may change in discontinuous steps only. Nevertheless the implementation of their method (i. e. starting from classical equations and discretizing "by hand") is awkward and would have to be adapted to every new problem treated. The aim to achieve a "final quantum theory" must be the discovery of generally applicable quantization conditions and "quantum dynamical equations of motion", directly addressing the problem of quantum transitions. The important hint from Heisenberg's paper towards this goal consists in the use of matrix multiplication rules for the quantum vectors (= amplitudes = matrix elements) introduced by Born and Jordan in BJ-1. But Heisenberg's continuous time behaviour, inconsistent with discontinuous spatial behaviour and discontinuous quantum transitions in general, will have to be eliminated.\\

In their paper BJ-2, Born and Jordan restrict the discussion of mechanical variables to a single degree of freedom; the generalization to arbitrary degrees of freedom is announced for a succeeding publication, which will follow shortly ( Born, Heisenberg, Jordan, Z. Phys.35,557,1926; rec. 16 Nov. 1925, referred to as {\bf  "BHJ"}). Similarly, in BJ-2 the treatment of the coupling to the electromagnetic field {\bf including field quantization} is derived for the coupling to a single electric dipole only. Field quantization however, is possible for all modes, due to the linear character of Maxwell's equations, allowing decomposition into linearly independent modes. \\

The key, which will open the path towards the solution of the quantum puzzle, is the general quantization condition. Born's quantization condition -  requiring the action variable  to change by integer multiples of $h$ only - is applied to the product of the canonically conjugated variables position $q$ and momentum $p$.  The classical variables are replaced  by their respective Fourier expansions and the classical Fourier coefficients are replaced by discontinuous matrix elements. Using matrix multiplication rules as suggested by Heisenberg, the classical product of $q$ and $p$ is thereby replaced by the matrix product $\tilde q \; \tilde p$. Requiring action to change by discrete steps of $h$ only, the general quantization condition is obtained:\\

\begin{equation}
\tilde p \;\tilde q - \tilde q\;\tilde p = \frac {h}{2\pi i} \tilde 1.
\end{equation}

The classical Hamiltonian equations retain their validity, if the classical variables are replaced by their associated matrices; using the general quantization condition, Born and Jordan show that the classical equations of motion may be transformed into new quantum mechanical "equations of motion" \\  
\begin{equation}
 \tilde {\dot q} = \frac { 2\pi i}{h} (\tilde H \;\tilde q - \tilde q\; \tilde H), \;\;\;\;\;\;\;\;\;\; \tilde {\dot p} = \frac {2\pi i}{h} (\tilde H\; \tilde p - \tilde p\; \tilde H),
\end{equation}
where $\tilde H$ is the matrix representing the Hamiltonian. \\

In their original form the "quantum theoretical equations of motion" are {\bf not differential equations}: their right hand sides are the commutators of two matrices; the resulting matrices are  the quantum mechanical equivalent of the classical variables $\dot q$ and $\dot p$, which, like any other physical variable, are represented by  matrices. For example:
The matrix element $\tilde {\dot  p}_{ba}$ {\bf is not} the time derivative of $ \tilde {p}_{ba}$, but is given by
\begin{equation}
 \tilde {\dot  p}_{ba} = \frac {2 \pi i}{h} \sum_{c}[\tilde H_{bc}\;\tilde p_{ca} - \tilde p_{bc}\; \tilde H_{ca}].
\end{equation}
This equation expresses the following physical content: The discontinuous transition from some initial state $|a\rangle $ to some final state $|b\rangle $ causes a particle to undergo a {\bf discontinuous momentum change}, which is expressed by the matrix element ${\tilde {\dot  p}}_{ba}$.
 In classical physics $ \dot q$ and $ \dot p$ are defined by quotients of  infinitesimally small intervals. According to Born's "Quantenmechanik" however, the basic quantum laws of nature  do not allow infinitesimally small intervals; the replacement of the infinitesimally small intervals of classical physics by the discontinuous matrix element  is the direct consequence of this fundamental principle\footnote {To avoid confusion: The indices $a$, $b$, $c$ may refer to a continuum of values, the {\bf transitions} from $a$ to $b$ or $b$ to $c$ are discontinuous; all transitions are quantized.}. As was already contained in BJ-1, the transition probability is proportional to $|{\tilde {\dot p}_{ba}|^2}$.
The particular form of the Hamiltonian matrix will determine which quantum transitions are possible.  The diagonal matrix elements are average values, for example:  $\tilde q_{aa}$ is the average position in state $|a\rangle $.\\
For a general physical variable $g$ and its matrix representation $\tilde g$, the corresponding "equation of motion" is shown to have identical form  
\begin{equation}
\tilde {\dot g} = \frac { 2\pi i}{h} (\tilde H \;\tilde g - \tilde g\; \tilde H). 
\end{equation}

Time  does no longer appear explicitly in the original form of these "equations of motion", but is contained implicitly in the requirement:\\
{\bf Transition probabilities per unit time are proportional to absolute squares of nondiagonal matrix elements}.\\
This requirement, the commutation relation, and the new "quantum equations of motion"  represent the basic quantum laws.\\

The final chapter of BJ-2 contains the coupling to the electromagnetic field, {\bf field quantization}, and the calculation of the radiation energy emitted in spontaneous transitions. Although the method used by Born and Jordan is unusual and will later be replaced by others, the essential elements of field quantization are described. Quantization of the action variable must affect electric and magnetic fields as well, which are represented by associated matrices.  Maxwell's equations are retained for matrices representing electric and magnetic fields. Born and Jordan conclude, that the method to solve the classical Maxwell equations may be carried over to the quantum problem for propagation in vacuum. The classical solution is obtained by decomposition of Maxwell's equations into infinitely many noninteracting modes of harmonic oscillators. The quantum behaviour of the simple harmonic oscillator may thus be applied to these modes and - since they are noninteracting -  to all. Furthermore, the interaction of a single mechanical degree of freedom to infinitely many noninteracting radiation modes may be carried out to lowest order perturbation theory, neglecting backcoupling effects. To calculate the radiation energy emitted, Born and Jordan determine the matrix equivalent of the Poynting vector for a radiating electric dipole; calculating the radiation energy emitted, the classical results are recovered for the averages.\\

\begin{subsection}
 { M. Born, W. Heisenberg, P. Jordan, Z. Phys.35,557,1926 }
\end{subsection}

The completion of the elementary Quantum Theory was a common effort by {\bf Born, Heisenberg, and Jordan (Zur Quantenmechanik. II.; Z. Phys.35,557,1926; rec. 16 Nov. 1925)}, which will be referred to as {\bf BHJ}.\\

Concerning the history of this important publication a few remarks are helpful. Heisenberg was absent from G\"{o}ttingen, when Born and Jordan worked on and completed their paper BJ-2, he had left in the middle of July to travel via Holland to England, accepting invitations by P. Ehrenfest to Leyden and R. H. Fowler to Cambridge. There Heisenberg described the ideas of his new paper and Fowler asked Heisenberg to provide a copy when available. 
In the middle of August Born informed Heisenberg about the progress achieved and the major results. Heisenberg sent a copy of the proofs  of his paper to Fowler towards the end of August, who passed it on to his young collaborator Paul Dirac. During the following months Dirac rederived the major results of Born and Jordan: Commutation relations and quantum equations of motion {\bf (P.A.M. Dirac, Proc.Roy.Soc. A 109, 642-653, 1925, rec. 7 Nov. 1925)}. 
Heisenberg returned to Copenhagen in the middle of September for a month long visit of Bohr's group. During his Copenhagen visit Heisenberg contributed to the paper BHJ by correspondence. He returned to G\"{o}ttingen in the middle of October, leaving little overlap with Born, who left G\"{o}ttingen on October 29 for a lecturing tour of the US. After an extensive stay at MIT, further lectures were given at other universities (Chicago, Wisconsin, Berkeley, Cal-Tech., Columbia); thereby the new message about quantum theory arrived in the US very quickly after its conception. Born returned to G\"{o}ttingen in March 1926. \\

A first draft of the paper BHJ had been prepared by Born and Jordan before Born's departure. The final version contains large sections rewritten by Heisenberg, who was still very much influenced by the Copenhagen concept. Heisenberg's letters to Pauli of 23 Oct and 16 Nov, 1925, testify of considerable differences in attitude and interpretation between the three authors, and these differences partly remained  afterwards  (later chapters will contain more details). Heisenberg left G\"{o}ttingen at the end of the winter semester (the end of February) 1926; he started his new position at Bohr's institute in Copenhagen on May 1st, 1926.\\

This paper by Born, Heisenberg, and Jordan ({\bf "BHJ"}) contains "almost everything" about the formal development of elementary quantum theory:\\
\noindent
- Treatment of systems with arbitrarily many degrees of freedom;\\
-  Perturbation theory for non-degenerate and a large class of degenerate systems;\\
-  The relation to the eigenvalue theory of Hermitian forms; discrete and continuous spectra;\\
-  Angular momentum algebra is developed;\\
-  General conservation laws (energy, momentum, angular momentum,  where applicable) are derived;\\
-  Quantization of the electromagnetic field, selection rules for radiative transitions, intensities of spectral lines; \\
-  Statistical treatment of black body radiation, quantum statistical derivation of Einstein‘s fluctuation formula. \\

\begin{subsection}
{ Pauli's solution of the hydrogen problem  }
\end{subsection}

In parallel to the conception of  BHJ, Wolfgang Pauli succeeded in solving the hydrogen problem using the new quantum theory of "matrix mechanics".  On 3 Nov. 1925 Heisenberg replies to a letter of Pauli (Pauli's letter is not conserved); Heisenberg expresses his joy about learning of Pauli's solution. The news is spreading quickly, even well before Pauli will eventually submit his paper ({\bf W. Pauli, Z. Phys. 36, 336, 1926, rec. 17 Jan 1926}). On 17 Nov 1925 Pauli sends a letter to Bohr with an extensive description of his results, including the Balmer formula and the results for the Stark effect. Bohr mentions Pauli's solution in publications in Nature (116, 845, 1925, appearing in print on 5 Dec. 1925) and in Naturwissenschaften (14, 1-10, 1926, in print on 1 Jan 1926) \\

Pauli solves the eigenvalue problem of the one body problem in an attractive $1/r$ potential using the algebraic methods of "matrix mechanics". He recognizes that the matrices corresponding to  the Hamiltonian, $1/2m \; {\bf p^2} - e^2/r$, the square of angular momentum ${\bf L^2 = (r \times p)^2}$, its  z-component $L_z$, and the quantum equivalent of the square of the "Laplace-Runge-Lenz vector", defined by \\

\begin{equation}
{\bf A}^2 = (\frac {1}{e^2m} \;{\bf  L \times p + r}/r)^2 \;,
\end{equation}
are all mutually commutative and can be diagonalized simultaneously. Using algebraic and combinatorial arguments Pauli obtains the complete solution. \\

\begin{subsection}
{ Remarks about differences in understanding between Born, Jordan, Pauli, Heisenberg and Bohr}
\end{subsection}

As mentioned before, Pauli and Heisenberg had both been students of Sommerfeld to obtain their doctorates in Munich, and both went to G\"{o}ttingen afterwards to work with Born. Born had recognized the difference between classical and  quantum dynamics (“action variables may change by integer multiples of h only”) already before Pauli became Born's "Assistent" in G\"{o}ttingen in 1921 (see his letter to Pauli of Dec. 19, 1919, mentioned earlier). Born suggested to Pauli to use Hamilton-Jacobi mechanics as method for perturbation theory (a common publication resulted: M. Born, W. Pauli, Z.Phys. 10, 137, 1922), the type of method which Born later used in his 1924 publication (introducing the term “Quantenmechanik”) and in his book "Vorlesungen \"{u}ber Atommechanik, 1. Band" (Nov. 1924), which were the starting points for the new quantum theory. But Pauli was dissatisfied with Born, he left G\"{o}ttingen after 6 months already to take a new post in Hamburg; afterwards Pauli typically made derogatory remarks about Born, criticizing Born's heavy mathematics! What Pauli (and even Heisenberg later on) failed to see, was that Born started from experimental observations (he called that the “documents of nature”), he then deduced physical arguments, and then he used the necessary mathematics to implement the physical ideas. \\

 Heisenberg's first collaboration with Born occurred during the winter semester 1922/23, when Sommerfeld was absent from Munich and had sent Heisenberg to work with Born during his absence. Born was impressed by 
Heisenberg\footnote {Concerning Born's attitude towards Pauli and Heisenberg, the letter to Einstein of 7 April 1923 contains: {\it "I had Heisenberg here during the winter.....equally as gifted as Pauli... but nicer and more pleasant"} } and offered Heisenberg to become his "Assistent" after the completion of Heisenberg's Doctorate in July 1923. Pauli's sceptical attitude towards Born influenced Heisenberg, who was looking more towards Bohr for physical concepts and ideas, considering Born too mathematical (see his letters to Pauli of 23 Oct and of 16 Nov 1925).\\

Heisenberg and Pauli were both strongly influenced by and had great admiration for Niels Bohr, who was generally considered to be the highest "quantum authority".
But whereas Pauli had very strong personal ideas 
already\footnote{Pauli's publications on the Compton effect (Z. Phys. 18, 272, 1923, and Z. Phys. 22, 261, 1924) relied on Einstein's light quanta, in disagreement with Bohr's continuous radiation. Pauli's letter of 12 Dec 1924 to Bohr (containing the  manuscript of the "Pauli principle", Z.Phys 31,765,1925) criticized Bohr's continuous orbits; Pauli expressed his support of Born's concept about discontinuous quantum transitions, characterizing stationary states only by quantum numbers.} 
Heisenberg still adhered rather strictly to the Copenhagen concept.
He saw the contributions of Born and Jordan mainly as mathematical techniques to treat the mechanical degree of freedom, without recognizing their full implications. 
This is particularly valid for field quantization, at that point of time still rejected by Bohr and his followers.  Field quantization was already part of BJ-2 (the calculations are mainly due to Jordan), and Heisenberg was not convinced, as his letters to Pauli testify. Even after the common paper BHJ of November 1925 (which includes an extensive chapter on field quantization, including  the derivation of Einstein's fluctuation formulas) Heisenberg did not accept field quantization.
 Heisenberg's letter to Pauli [105] of 16 Nov 1925 (sending the final version of BHJ) contains quite strange ideas about classical radiation. Pauli replied quickly; the letter is not preserved, but according to Heisenberg's reply [108] of 24 Nov 1925, it must have contained the same information as Pauli's letter to Bohr [106] of 17 Nov 1925, mentioned above. Heisenberg revised his position somewhat, but he retained Bohr's idea about continuous radiation, as stated explicitly in a presentation to the German Mathematical Society on 19 December 1925 
(published in Math. Ann. 95, 683-705, 1926).\footnote{Field quantization will obtain wider recognition only after Dirac's stay in G\"{o}ttingen from May to September 1926, from where he moved on to Copenhagen, and Dirac's publication on field quantization  {\bf (Proc. R. Soc. Lond. A 114, 243-265, 1927)} carried out in Copenhagen  afterwards.}\\

Concerning the question  "what are the essential results of the new quantum theory?", essential differences in understanding exist between Born and Jordan on one side, and Heisenberg and Bohr on the other. For Heisenberg the diagonalization of the Hamiltonian matrix to obtain Bohr's stationary states represents the essential result, in his mind that amounts to "integration of the equation of motion", as stated in his letters to Pauli. Stationary states correspond to static properties, there is no quantum dynamics contained and the transition processes are still open. As Born and Jordan had stated, energies themselves are not observable; observable quantities result from transitions; in radiative transitions, for example, only the emitted radiation  is observable, providing information about energy differences. For Born and Jordan {\bf "quantum dynamics"} of  transition processes,  in particular the determination of transition probabilities, constituted the essential aim. The diagonalization of the atomic Hamiltonian on one side and the quantization of the free electromagnetic field on the other are only a preliminary step towards the determination of transition probabilities, caused by the mutual interaction between field and electrons. \\

Essential differences in physical understanding will also be apparent concerning the origin of quantum uncertainties, which will be discussed extensively in a later chapter.\\

\begin{subsection}
{ Brief summary of the New Quantum Theory }
\end{subsection}

\noindent
 Let us summarize the new message contained in "matrix mechanics":\\
All physical variables have lost their classical significance, being replaced by an associated matrix; particle position  or any other variable do no longer have precise values, only statistical statements are possible. Time as well has lost its traditional meaning; just as position can no longer be exactly assigned to a particle, the exact time of a quantum transition can not be given. The general quantum conditions take the form of commutation relations; the "quantum theoretical equations of motion" have become relations between matrices; for example the time derivative of momentum in classical physics is replaced by a matrix, which, via the new "quantum mechanical equation of motion", is related to the commutator of the Hamiltonian matrix with the matrix representing the momentum.\\

\noindent
Physical content and mathematical form of the new Quantum Theory were so dramatically different from all traditional concepts in classical physics, that it is no wonder, that the new basic laws shocked the established scientific community, and considerable resistance was widespread. Determinism and continuity, mathematically expressed via differential equations, had been the foundation of classical physics, and it is maybe not surprising that a return to seemingly familiar concepts appeared soon afterwards. \\

\begin{section}
 {Continuous Representations of the New Quantum Laws}
\end{section}

\begin{subsection}
 {K. Lanczos: field theoretical representations}
\end{subsection}

{\bf Kornel Lanczos ( Z.Phys. 35, 812, 1926, rec. 22 Dec. 1925)} was the first to point out, that the discontinuous form of the new fundamental  laws of Quantum Mechanics could mathematically be represented by field theoretical methods, relying on functions of continuous variables. The algebraic eigenvalue equations of Born, Heisenberg, Jordan 
(for example of the Hamiltonian matrix)
can be represented by integral equations such that the eigenvalues of the integral operator yield the inverse of the eigenvalues of the algebraic equation. Similarly differential equations might be used, yielding the same eigenvalues as the algebraic equations.  Lanczos stressed, however, that field theoretical representations do not imply physical continuity and Lanczos' publication contains a warning: If all physical processes are indeed discontinuous and all physically relevant quantities are contained in discrete matrix elements, then the field theoretical representations necessarily contain an additional arbitrariness. Infinitely many different representations using different variables are possible yielding the same eigenvalues; the functions of continuous variables then are nothing more than mathematical auxiliary functions, which may be used to calculate the physical significance contained in averages and transition probabilities.\\

\begin{subsection}
{ Linear Hermitian operators and "time representation"}
\end{subsection}

The first partially continuous representation was proposed  by  {\bf Max Born and Norbert Wiener (Z.Phys. 36, (1926), 174-187, rec. 05 Jan 1926)}, when they introduce linear Hermitian operators acting in an infinite dimensional vector space (later called "Hilbert space"), replacing matrices.  The previous matrix elements may be written as\\

\begin{equation}
\tilde p_{ba} = \langle  b|\hat  p|a\rangle , 
\end{equation}

where $\hat p$ is an operator. The basic laws are formulated for Hermitian operators instead of matrices;
for any pair of canonically conjugated variables $q$ and $p$ the general quantization condition is contained in the commutation relation for  corresponding Hermitian operators

\begin{equation}
\hat p \hat q - \hat q \hat p = \frac {h}{2\pi i} = \frac {\hbar}{i} .
\end{equation}

Similarly the "equations of motion" are formulated for operators 

\begin{equation}
\hat {\dot q} = \frac {i}{\hbar} (\hat H \hat q - \hat q \hat H), \\ \hat {\dot p} = \frac {i}{\hbar} (\hat H \hat p - \hat p \hat H).
\end{equation}

Referring to the Hamilton-Jacobi formalism of classical mechanics, Born and Wiener remark that "time" and "energy" may be considered canonically conjugated variables, just as any other pair of generalized coordinates and momenta. "Time" may take the role of generalized coordinate and the negative "energy" will be the canonically conjugated momentum; Born and Wiener conclude that the commutation relation for time and energy will have to be fulfilled.\footnote{ As will be discussed later in the chapter on "Quantum Uncertainties", commutation relations require the uncertainties of canonically conjugated variables to be connected; for time and energy this relation couples lifetime and energy uncertainties.}
 Remember that the origin of Born's quantization condition resulted from the same principle applied to the canonically conjugated variables "action" and the associated dimensionless "angle variable". \\

In the previous publications BJ-1, BJ-2, and BHJ "time" had disappeared as explicit variable, being contained implicitly only in "transition probability per unit time". Now Born and Wiener introduce a special representation, where time $t$  is taken to be a continuous variable ("time representation").  To fulfill the commutation relation for time $t$ and energy $\hat E$ 

\begin{equation}
[\hat E t - t \hat E] = - \frac {\hbar}{i},
\end{equation}
the {\bf energy becomes the operator: $\hat E = -  \hbar /i  \cdot  d /d t$}. A differential equation with respect to time is obtained

\begin{equation}
H |u(t)\rangle  = - \frac {\hbar}{i} \frac{d}{dt}|u(t)\rangle .
\end{equation}

The variation of $|u(t)\rangle $  with respect to time $t$ does not describe the continuous variation of the physical state of the system;
$|u(t)\rangle $  is a mathematical auxiliary object, which may be used to calculate matrix elements and the relevant averages and transition probabilities.\\

The "equations of motion" for the operators can now be taken to be differential equations with respect to the continuous variable $t$; formal integration yields

\begin{equation}
\hat q(t)  = e^{\frac {i}{\hbar}\hat H t} \; \hat q \; e^{- \frac {i}{\hbar}\hat H t}, \\ \hat p(t)  = e^{\frac {i}{\hbar}\hat H t} \; \hat p \; e^{- \frac {i}{\hbar}\hat H t}, 
\end{equation}

and $|u(t)\rangle $ takes the form

\begin{equation}
|u(t)\rangle  =  e^{- \frac {i}{\hbar}\hat H t}|u\rangle .
\end{equation}
\\

\begin{section}
{ "Wave Mechanics"}
\end{section}

During the first half of 1926, Erwin Schr\"{o}dinger published a series of 5 papers in quick succession ( Ann. Phys. 79, 361-376, 1926; Ann. Phys. 79, 489-527, 1926; Ann. Phys. 79, 734-756 1926; Ann. Phys. 80, 437-490, 1926; Ann. Phys. 81, 109-139, 1926), which contained the alternative formulation of "wave mechanics". Schr\"{o}dinger refers to the thesis of {\bf Louis de Broglie}, presented in early 1924 ({\bf "Recherches sur la Theorie des Quanta", Ann. de Physique 3, 22-128, Jan. 1925}), which Einstein had already  mentioned, when he applied Bose statistics to discuss ideal gases (Sitz. Berlin Ak. d. Wiss. 10-07-1924, and  08-01-1925).  Before describing Schr\"{o}dinger's publications in more detail, a brief discussion of de Broglie's ideas will be given.\\

\begin{subsection}
{ De Broglie: Particles and Associated Phase-Waves}
\end{subsection}

De Broglie's starting point was Einstein's theory of special relativity (Ann. Phys. 17, 891-921, 1905; and Ann. Phys. 18, 639-641, 1905), the equivalence of energy and mass and Einstein's photon concept of quantized radiation. De Broglie suggested that photons are particles with extremely small mass, which he called "light atoms" ({\it "atomes de lumiere"}).  He associated an oscillation process of frequency $\nu$ to these "light atoms" via the relation $mc^2 = h\nu$, with $m =m_0 (1-v^2/c^2)^{-1/2}$, and $m_0$ being the rest mass. The velocity $v$ of the "light atom" should be extremely close to the "maximal velocity" $c$, such that in this extreme relativistic regime all possible frequencies may be reached while the velocity changed by immeasurably small amounts only. The rest mass $m_0$ of the "light atoms" should be so small, that in the entire experimentally accessible region the mass $m$ should still remain to be immeasurably small.\\

De Broglie associated an equivalent oscillation process, which he called "phase waves",  to particles with finite mass, in particular to electrons. These "phase waves" should extend over all of space, energy and momentum of "light atoms" and of electrons, however, should still be concentrated in extremely small regions of space. In this sense, de Broglie retained the particle character of "light atoms" and electrons. \\
For electronic velocities $v$ (for example in x-direction and measured in some "reference system") the associated oscillation process $sin [2 \pi \nu (t - vx/c^2)]$ corresponded to a phase velocity $c^2/v$, which, however, should not represent a physical process; the physical velocity of the particle carrying its energy and momentum should be given by the group velocity, which, in all cases, would be smaller that $c$.\\
Finally, the oscillation process connected wavelength $\lambda$, velocity $v$, and momentum $p$ via Planck's constant: $h/\lambda = mv = p$.
\\

Based on these hypotheses de Broglie had given a new interpretation for the quantization conditions of the "old quantum theory": The length of the periodic orbit of stationary states  should be an integer multiple of the wavelength $\lambda$. \\

\begin{subsection}
{Schr\"{o}dinger's "Position Representation" }
\end{subsection}

Already before the series of five publications introducing "wave-mechanics" Schr\"{o}dinger announced his intention to take de Broglie's wave concept even more seriously than de Broglie himself. On 15 Dec. 1925 his paper "Zur Einsteinschen Gastheorie"  ({\bf Phys. Z. 27, 95-101, 1926}) was received for publication, in which Schr\"{o}dinger took issue with Einstein's application of Bose statistics to ideal gases. Schr\"{o}dinger rejected the new statistics for particles; as far as Einstein's particle concept of quantized radiation is concerned, Schr\"{o}dinger insisted that an oscillating wave picture should be maintained. The possible excitations of these oscillations should be required to occur in integer multiples of $h\nu$ only, which effectively amounted to applying Bose statistics to wave excitations. Bose's quantized phase space volume in terms of $h^3$ was replaced by an equivalent density of allowed waves per frequency interval (imposed by appropriate boundary conditions).  Schr\"{o}dinger announced to take seriously the de Broglie concept for particles with finite mass as well. Moving particles should be viewed as {\it "a kind of wave crest on top of a world background of wave radiation"}. Schr\"{o}dinger effectively left the Bose-Einstein mathematics invariant, just the interpretation was 
different. Whereas Einstein in 1905 had claimed that electromagnetic radiation - just as matter - consisted of elementary objects having particle character, Schr\"{o}dinger took the opposite direction; particles with finite mass and radiation should "really" be taken to be waves. \\

Although Schr\"{o}dinger will refer to de Broglie and waves throughout the series of 5 publication from January to June 1926, his mathematical treatment does not contain any relation to de Broglie's. Schr\"{o}dinger's first publication of the series {\bf ( Ann. Phys. 79, 361-376; rec. 27-01-1926)} does not describe the attempt to derive a wave equation, instead an eigenvalue equation is proposed treating the hydrogen problem, which Pauli had solved shortly before using the new "matrix-mechanics". Schr\"{o}dinger claims to derive the equation \\

\begin{equation}
{\bf \Delta} \psi + \frac {2m}{K^2} (E + e^2/r) \psi = 0
\end{equation}

from the Hamilton-Jacobi partial differential equations. K is to be identified as $\hbar = \frac{h}{2\pi}$ in order to reproduce the Balmer series of hydrogen. Schr\"{o}dinger's solution of the eigenvalue equation is reproduced in many text books on Quantum Mechanics.\\

The procedure from the Hamilton-Jacobi equations $H(q, \partial S/\partial q) = E$ to the eigenvalue equation appears to be arbitrary. Schr\"{o}dinger first replaces the action $S$ by $K ln\psi$, then , instead of solving the resulting differential equations, he introduces a variational procedure

\begin{equation}
\delta \int d^3{\bf r} \; [(\partial \psi /\partial x))^2 + (\partial \psi /\partial y))^2  + (\partial \psi /\partial z ))^2 \; - \frac {2m}{K^2} (E + e^2/r) (\psi)^2 ] = 0
\end{equation}

to arrive at the eigenvalue equation above.  If the procedure seems more plausible as an attempt to work backwards, starting from the eigenvalue equation towards the Hamilton Jacobi differential equations, the question remains open, how Schr\"{o}dinger obtained the equation ${\bf \Delta} \psi + \frac {2m}{K^2} (E + e^2/r) \psi = 0$.  Schr\"{o}dinger himself does not give any indications. \\

Effectively Schr\"{o}dinger  has introduced the "position representation",  where position ${\bf r}$ is  taken to be continuous variable ${\bf r}$. The quantization condition, $[{\bf \hat p r - r \hat p}] = \hbar/i$  is fulfilled, the operator ${\bf \hat p}$  is replaced by the gradient with respect to position:  ${\bf \hat p} = \frac {\hbar}{i} {\bf \nabla_r}$. Six weeks later, in his third paper, Schr\"{o}dinger ({\bf Ann. Phys. 79, 734-756 1926, rec. 18 March 1926)} will make this connection explicitly. Lanczos had pointed out that a differential equation could be formulated containing the quantization condition implicitly, and the transition from the commutation relation $[{\bf \hat p r - r \hat p}] = \hbar/i$  to the 
continuous "{\bf r}" representation" and ${\bf \hat p} = \frac {\hbar}{i} {\bf \nabla_r}$  appears logical.\\

In his second publication of the series ({\bf Ann. Phys. 79, 489-527, 1926, rec. 23 Feb 1926}) Schr\"{o}dinger, besides making a first attempt to propose a wave equation,  discards the derivation of his first publication, in particular the substitution $S$ by $K ln\psi$, and proposes as different method relying on the analogy contained in the Hamiltonian variational principle between optics, based on Fermat's principle, and the principle of least action of mechanics. 
The analogy involves geometric optics not wave optics and serves to replace the "derivation" of the previous publication to arrive at the  eigenvalue equation\footnote{Schr\"{o}dinger's use of dimensions is somewhat confusing, he leaves out the mass $m$ in this equation. In the following $m$ will be included}
 ${\bf \Delta} \psi + \frac {2m}{\hbar^2} (E - V) \psi = 0$. \\

\noindent
The question of obtaining a proper wave equation still had to be addressed. Schr\"{o}dinger assumes harmonic behaviour in time ($\psi \sim e^{\frac {i}{\hbar} Et}$)  with frequency $E/\hbar$    and aiming for a wave equation of second derivative with respect to time, he replaces:\\
$(E-V) \; \psi = - \; \hbar^2\; \frac {E-V}{E^2}  \frac {\partial ^2}{\partial t^2} \psi $,
to arrive at the "wave equation"

\begin{equation}
{\bf \Delta} \psi - \frac {2 m (E - V)}{E^2} \frac {\partial ^2 }{\partial t^2}\psi =  0.
\end{equation}
But this equation cannot serve as a proper wave equation, since it contains the energy explicitly, restricting the possible solutions to harmonic behaviour. The "wave equation" is equivalent to the time independent eigenvalue equation. \\

Of particular interest is Schr\"{o}dinger's first reference to the quantum theory of Born, Heisenberg, and Jordan and the description of Schr\"{o}dinger's attitude towards the new mode of thought it contains. Schr\"{o}dinger criticises the elimination of  continuity in space and time; he argues that - from the philosophical point of view - this constitutes total capitulation , which he refuses: {\it "Denn wir k\"{o}nnen die Denkformen nicht wirklich \"{a}ndern und was wir innerhalb derselben nicht verstehen k\"{o}nnen, das k\"{o}nnen wir \"{u}berhaupt nicht verstehen."  ("We are not really able to change the modes of thought and if we cannot understand within these modes of thought, then we cannot understand at all.")} A discussion contrasting the attitudes of Einstein, Bohr, Born, and Schr\"{o}dinger will be given in a later chapter.\\

Schr\"{o}dinger's third publication of the series ({\bf Ann. Phys. 79, 734-756 1926, rec. 18 March 1926)} contains the explicit connection to "matrix mechanics" and the transition from the commutation relation $[{\bf \hat p r - r \hat p}] = \hbar/i$  to the 
continuous "{\bf r} representation" and ${\bf \hat p} = \frac {\hbar}{i} {\bf \nabla_r}$. But Schr\"{o}dinger discards a full equivalence of the two approaches,  insisting on the physical significance of  wave functions, which he sees as the essential element; "wave mechanics" should be understood as an extension and part of classical field theories.  Concerning "matrix mechanics" Schr\"{o}dinger reaffirms  his disagreement about physical content and mathematical formulation; he states  that he felt deterred, if not to say repelled, by the apparently "very difficult methods of transcendental algebra and the lack of illustrative clarity" ({\it "ich f\"{u}hlte mich durch die mir sehr schwierig scheinenden Methoden der transzendenten Algebra und durch den Mangel an Anschaulichkeit abgeschreckt, um nicht zu sagen abgestossen"}). \\

Schr\"{o}dinger's forth publication of the series ({\bf Ann. Phys. 79, 734-756 1926, rec. 10 May 1926)} contains the "wave mechanical" version of time independent perturbation theory (equivalent to "matrix mechanical" perturbation theory of "BHJ") and its application to the Stark effect of the Balmer series, reproducing the results Pauli had obtained in his "matrix solution" of the hydrogen problem. \\

In his  5'th publication of the series {\bf (Ann. Phys. 81, 109-139, 1926; rec. 21 June 1926)} Schr\"{o}dinger recognizes that the time dependent "wave equation" of the second publication cannot be used as general wave equation, since energy $E$ is contained explicitly.  \\
Schr\"{o}dinger makes several further guesses at wave equations, which originate from the application of the operator $({\bf \Delta} - \frac {2m}{\hbar ^2}  V)$  to the eigenvalue equation for a second time,  leading to an equation containing forth order derivatives with respect to position.
Again simple harmonic behaviour with frequency $\nu = E/\hbar$ is provisionally assumed (equivalent to: $(\frac {E}{\hbar})^2  \psi = -\frac {\partial ^2 }{\partial t^2}\psi $)  leading to 

\begin{equation}
({\bf \Delta} - \frac {2m}{\hbar ^2}  V)^2 \psi = - \frac {4m^2}{\hbar ^2} \frac {\partial ^2 }{\partial t^2}\psi,
\end{equation}
which is then proposed to be the {\it "general wave equation for the field scalar $\psi$"}.\\
Two more versions follow, which are obtained from factorizing the previous time independent equation: 
$({\bf \Delta} - \frac {2m}{\hbar ^2}  V \pm \frac {2m}{\hbar ^2} E) ({\bf \Delta} - \frac {2m}{\hbar ^2}  V \mp \frac {2m}{\hbar ^2} E)\psi = 0$. At this point again provisionally assuming behaviour in time $ \psi  \sim  e^{i/\hbar \cdot Et}$, the equations 

\begin{equation}
({\bf \Delta} - \frac {2m}{\hbar ^2}  V) \psi = \pm \frac {2m i}{\hbar} \frac {\partial  }{\partial t}\psi
\end{equation}
are obtained. Schr\"{o}dinger proposes that either of these equations might serve as generally valid wave equation. \\

We recognize that the "time representation" introduced by Born and Wiener combined with Schr\"{o}dinger's "position representation" yields the equation with the minus sign, now called the time dependent Schr\"{o}dinger equation.\\

\begin{subsection}
{ Max Born: The probabilistic significance of wave functions }
\end{subsection}

Very quickly after
Schr\"{o}dinger submitted his version of "wave mechanics", {\bf Max Born  (Z. Phys. 37, (1926), 863-867, rec. 25 June 1926)} established the relation 
between "wave functions", matrix elements and probabilities. 
Born discussed the scattering of an electron by an atom
and he used Schr\"{o}dinger's "position representation" to calculate
the relevant matrix elements. Born accepted Schr\"{o}dinger's "position 
representation" as mathematical tool to calculate matrix elements, 
but he insisted that Schr\"{o}dinger's physical interpretation was
incorrect;  Born writes:\\
{\it  Man bekommt keine Antwort auf die Frage,"wie ist der Zustand
 nach dem Zusammenstoss", sondern nur auf die Frage,"wie wahrscheinlich ist ein
 vorgegebener  Effekt des Zusammenstosses"....Vom Standpunkt unserer
 Quantenmechanik  gibt es keine Gr\"{o}sse, die im Einzelfall den Effekt des
 Stosses festlegt."}\\
({\it "We do not get an answer to the question, "what is the state after the collision?", 
but only to the question, " how probable is a given result of the collision?"...
 based on the principles of our Quantum Mechanics there exists no quantity, 
which determines the result of the collision for the individual elementary process."})\\

Born describes the process of an incoming electron with momentum ${\bf p}_i$ and kinetic energy $\epsilon_i = p_i^2/2m$ being scattered by an atom having initial energy $W_i$. The observable quantities are the incoming and outgoing momenta ${\bf p}_i$ and ${\bf p}_f$. In the spirit of first order perturbation theory the scattering process is taken to result from a single elementary quantum transition induced by the interaction $V_{int}$ between electron and atom to a final electronic state with momentum ${\bf p}_f$ and kinetic energy $\epsilon_f = p_f^2/2m$, the atom undergoing a transition to a final state of energy $W_f$. Born uses Schr\"{o}dinger's position representation to describe the scattering process; the electron is represented by plane wave functions, the incoming state as $\psi_i = sin( 2\pi /\lambda_i\; z)$, the outgoing state as $\psi_f = sin \; (2\pi /\lambda_f \;(\alpha x + \beta y+\gamma z)) $. Wavelengths and electron momenta are related by $p = h/\lambda$. Total energy conservation requires the outgoing wavelength to be given by $(h/\lambda_f)^2/2m = W_i - W_f+ \epsilon_i$. The transition probabilities are determined to be proportional to $|\langle \Psi_f,\psi_f|V_{int}|\Psi_i,\psi_i\rangle |^2$, where $\Psi_i$ and $\Psi_f$ are intial and final atomic states.\\

Let us summarize the message from Born's "user manual" for Schr\"{o}dinger wave functions: Although Schr\"{o}dinger's position representation may be used to correctly calculate physically relevant matrix elements, the  solutions to the Schr\"{o}dinger differential equations do not have direct physical significance; \\
{\bf the "wave-functions" $\psi ({\bf r},t)$ are nothing more than mathematical auxiliary functions to calculate averages and probabilities}. \\

\begin{subsection}
{ Brief summary }
\end{subsection}

Schr\"{o}dinger's  "position representation" and the resulting wave functions reduced the unfamiliar form of algebraic equations derived by Born, Heisenberg, and Jordan  to the familiar territory of differential equations. This was widely considered to be a major advantage and Schr\"{o}dinger's wave equations became the method of choice for elementary applications of the new quantum theory. But the technical advantage of what Schr\"{o}dinger called "wave mechanics" became a major obstacle for the understanding of quantum mechanics. Whereas the discrete form of "matrix-mechanics" directly indicated the fundamental property of discontinuous quantum physics, Schr\"{o}dinger's  differential equations and the resulting wave functions suggested continuity and determinism.  Schr\"{o}dinger himself interpreted the "wave functions" to have direct physical significance, attributing "true wave character" to single particles.  Schr\"{o}dinger's aim had been to go back to the classical concepts of continuity in space and time, eliminating discontinuous and statistical quantum transitions and returning to continuity and determinism. This interpretation was adopted by many contemporaries.\\

 Schr\"{o}dinger's differential equations made quantum theory more accessible for calculational purposes, but the understanding of the principles of quantum physics suffered. This is particularly true for scattering experiments and the attribution of "wave properties" to particles, which directly touches the problem Einstein had pointed to in 1905, when he claimed that electromagnetic radiation consisted of elementary objects having particle properties. 
\\

\begin{section}
{ The solution to Einstein's problem: How to connect particle properties and "wave" phenomena}
\end{section}

Already before a valid quantum theory was available, {\bf William Duane (Proc. Nat. Ac. Sci. 9, 158-164, 1923)} suggested that phenomena observed in light scattering and X-ray scattering, which had been interpreted to result from "constructive and destructive interference" of waves, could be understood as a pure quantum phenomenon based exclusivly on particle characteristics of photons.\\

 Let us recall how the "spectral decomposition" of light to determine the Planck spectrum had been performed experimentally. Light was scattered off artificially grated surfaces in grazing incidence. The observation of special reflection maxima and minima was interpreted as a measurement of the "wavelength" of the "incident waves", due to "constructive and destructive interference" when scattered from the periodic grating. Using the relation $\nu = c/\lambda$  the "frequency" was determined. Radiation reflected from thin plates was interpreted in similar fashion. \\
In 1912 Max von Laue (Sitzungsberichte der Math.-Phys. Akademie der Wissenschaften (M\"{u}nchen) 1912; 303) discovered X-ray diffraction by crystals, which was generally considered to prove that X-rays were just another form of electromagnetic radiation of smaller wavelengths. The occurrence of diffraction peaks was interpreted to result from "constructive interference" of waves reflected from parallel lattice planes.\\

Wave characteristics of light and X-rays - such as wavelengths and frequencies - are not observable directly; the assumed "wave properties" result from indirect interpretations invented to "explain" the observed scattering maxima and minima. \\

Duane suggested that these phenomena are due to a quantum condition imposed on the momentum transfers of the scattered particles. Einstein's particle concept of light was taken to be correct; photon characteristics were energies $\epsilon$ and momenta $\epsilon/c$; no wave properties such as wavelengths or frequencies were required. Duane's arguments were based on a simple dimensional analysis: The selection of special momentum transfers $\Delta p$ should result from a quantum condition involving Planck's constant $h$. To arrive at an equation, a characteristic lengthscale had to be introduced, which - for diffraction phenomena - must be the periodicity length $d$ of the grating or the crystal, leading to $d \cdot \Delta p = n h$. Quantum theory (and experiments when the appropriate experimental techniques became available) proved Duane to be right!\\

To avoid the misconception that wave functions and "dual properties" - particle like and wave like -  of photons (or other particles such as electrons or neutrons) are necessary for the understanding of so called "interference phenomena", mathematical representations may be used, which describe the observable quantities by real numbers. Following the development of the new quantum theory by Born, Heisenberg, Jordan, and Wiener, different representations of the fundamental laws were proposed by Pauli, which are particularly useful to describe scattering processes.\\

On 14 Jan 1926 Heisenberg sent copies of the Born-Wiener paper and of Lanczos paper to Pauli, who quickly  understands the connection between commutation relations and continuous representations. Born and Wiener had used the time-energy commutation relation to introduce the "time representation", where time $t$ is continuous variable and the energy becomes the operator  $- \hbar /i  \cdot  d /d t $. 
Pauli's reply to Heisenberg of 31 Jan 1926 mentions the possibility of an equivalent "energy representation", where energy remains to be real variable $E$, and time becomes the 
operator  $ \hbar /i  \cdot  d /d E $.\footnote{Pauli will revert his position about this point later. In his book "Die allgemeinen Prinzipien der Wellenmechanik", Springer Verlag 1933, Pauli concluded that the existence of discrete spectra is incompatible with the energy-time commutation relation. Pauli erroneously claimed that time $t$ always has to be treated as an ordinary number. The following chapter on "Quantum Uncertainties" will contain a detailed discussion of this point.}  
Pauli's letter of 19 Oct 1926 to Heisenberg contains the definition of the "momentum representation"; momentum is represented by the real variable ${\bf p}$, and the commutation relation requires position to be represented by the gradient with respect to 
momentum: $\hat{\bf r} = - \hbar /i \;\bf {\nabla_p}$. \\
The momentum representation is particularly useful to describe the solution to Einstein's problem; in scattering experiments, initial and final momenta are the measured quantities and their representation by real variables makes the connection between experimental observations and physical content clearer.\\

Consider purely elastic scattering by a potential $V({\bf r})$. 
For $\bf p$ real variable and  $\hat{\bf r} = - \hbar /i \;\bf {\nabla_p}$, the transition for a particle with initial momentum ${\bf p}$ to a state of final momentum  ${\bf p} + \hbar {\bf q}$ may  mathematically be expressed by\\

\begin{equation}
e^{- i \bf q  \hat r} |{\bf p}\rangle  = |{\bf p} + \hbar {\bf q}\rangle  \;.
\end{equation}

For a scattering potential $ V({\bf \hat r}) =  \int_{\bf q} \tilde V({\bf q}) \; e^{- i \bf q  \hat r}  $ the probability for the particle to be scattered with  momentum transfer $\Delta {\bf  p} = \hbar {\bf q}$ will be proportional to $|\tilde V({\bf q})|^2$. For a structure which is translationally invariant if displaced by a a set of real space vectors ${\bf d_i}$, finite Fourier components are restricted to ${\bf q \cdot d_i} = 2\pi n$. The possible  momentum transfers have to fulfill the condition  $\Delta {\bf p \cdot d_i} = nh$, as suggested by Duane in 1923.\\
The selection of the allowed momentum transfers is a direct consequence of Born's quantization condition; all quantum transitions require the action variable to change by integer multiples of $h$.  For the scattering processes discussed here, the product of a vector ${\bf d}$, representing  a discrete translational symmetry, and the allowed momentum transfers $\Delta {\bf p}$ correspond to the change in action variables of the transitions, which - as required - are integer multiples of $h$.\\

 The second condition to be fulfilled is energy conservation. The special momentum transfers depend on the type of particle to be scattered only in so far as its relation between energy and momentum is concerned. This "dispersion relation" will be different for photons, neutrons, or electrons; but the possible momentum transfers are solely imposed by the periodicity of the scattering potential;  transferred momenta will have to be  identical for all types of particles: photons, electrons, neutrons,.....\\

Thermal radiation consists of statistically emitted particles without any wave 
characteristics.
The "spectral decomposition" obtained in a grating spectrum constitutes a decomposition according to different photon momenta. Incoming photons scattered  elastically obtain the same momentum transfer and different initial momenta are reflected under different angles. "Constructive and destructive interference" does not correspond to any real physical process; experimentally observable are 
particles\footnote  {To actually resolve the scattering of individual photons is experimentally difficult (possible today but not at the time). Nowadays reduced intensity and appropriate detectors make the observation of individual photon scattering possible. The electromagnetic waves described by Maxwell Theory (e. g. radiated by a transmitter) require emission of a macroscopic number of photons with correlated polarisations; classical wave properties are connected with the polarisation structure, which is imposed by the source (the transmitter). Single photons, however, do not have any wave properties, they are simply particles.} 
 scattered according to the statistical laws of quantum theory. \\

The results above may also be applied to other geometries defining the scatterer. Fourier decomposition and restriction to lowest order perturbation theory - equivalent to the occurrence of a single elementary quantum transition - allows the application to each Fourier component separately. All types of diffraction phenomena -  including the "double or multiple slit" experiments - are covered. \\

Let us recall the conditions imposed on the scattering contributions which had been  attributed to "interference phenomena": The transitions must be purely elastic, not inducing any change in quantum numbers characterizing the scatterer. These purely elastic events occur with finite probability only, scattering events accompanied by transitions within the scatterer contribute with finite probability as well. E. g. scattering contributions from  a crystal will contain processes which are accompanied by localized transitions in atoms; these events contribute finite probabilities to a wide range of possible momentum transfers, generating a probability pattern containing a more or less uniform background, in addition to the Bragg peaks resulting from elastic contributions. A detailed description of scattering processes will be given in the appendix.\\

\begin{section}
{Quantum uncertainties}
\end{section}

Already before a valid quantum theory had been developed, Bose's "quantization of phase space" (i.e.  particles can no longer be characterized by precise values of position and momentum, only the association with a finite elementary volume of size $h^3$ may be
 specified)
 indicated that quantization is connected with quantum uncertainties. When  Born introduced "Quantenmechanik" in 1924, quantum uncertainties of physical variables were contained as necessary elements; quantization of the action variable for all elementary transitions implied that physical variables could only be associated with averages over corresponding action intervals of size $h$. \\

\noindent
{\bf Heisenberg (Z.Phys. 43, 172, 1927, rec 23 Mar 1927)}
drew special attention to uncertainty relations of canonically conjugated variables and their consequencies for experimental observations.
The development of Heisenberg's ideas leading to this paper may be traced in the frequent exchange of letter's with Pauli.
Of particular interest is Pauli's letter of  19 Oct 1926 with reference to Born's publication (Z. Phys. 37, 863, 1926; rec. 25 June 1926) on scattering, transition probabilities and their relation to matrix elements. Born had used  Schr\"{o}dinger's position representation\footnote {Heisenberg had criticized Born's publication and the use of Schr\"{o}dinger's wave functions (letter to Pauli of 28 July 1926 [142]).}, Pauli pointed out that a "momentum representation" might be defined as well.
Furthermore Pauli remarks that the general commutation relation for canonically conjugated variables ($pq - qp = \hbar/i$)  implies that either $p$ or $q$ can be taken to be arbitrarily precise, but only at the expense of increasing uncertainty of the  canonically conjugated variable. Pauli considers this to be a "dark point", which will have to be cleared up. Heisenberg reacts (letter of 28 Oct 1926): {\it "I am very enthusiastic, because one can understand the physical significance of Born's formalism much better"}.  \\

Heisenberg's intention is expressed in the title of his paper: {\it "\"{U}ber den anschaulichen Inhalt der quantentheoretischen Kinematik und Mechanik"} ("On the illustrative content of quantum theoretical kinematics and mechanics");  he wants to give a {\it "illustrative or graphically appealing"} understanding ({\it "anschauliches Verst\"{a}ndnis"}) of the commutation relation and the impossibility for arbitrary precision of canonically conjugated variables.  This "illustrative understanding", however,  uses illustrations and concepts borrowed from classical physics, which are not suited to "explain" quantum behaviour. Heisenberg's essential argument for commutation relations and quantum uncertainties relies on the erroneous assumption, that measurements necessarily introduce disturbances ({\it "St\"{o}rungen"}) to the system to be measured; {\bf quantum uncertainties} of physical variables - {\bf according to Heisenberg} - were to be {\bf caused by measurements}.\\

Taking the measurement of position, for example, the line of thought goes as follows: If the position of particle "A" is to be measured,  light or X-rays or other particles "B" have to be scattered off particle "A". Heisenberg argues that this scattering process will  necessarily disturb, i.e. change, the original state of particle "A", in particular it will transfer momentum to the particle "A". This disturbance caused by the measuring process of position should - according to Heisenberg - be responsible for a "momentum uncertainty" of particle "A". 
To obtain a semi-quantitative estimate, Heisenberg invokes an analogy with the optical microscope, where the classical resolution is limited to lengths of order wavelength $\lambda$, and extends this reasoning to the "$\gamma$-ray-microscope". The coincidence measurements of {\bf W. Bothe and H. Geiger (Z. Phys. 32, 639, 1925)}  had convinced Heisenberg of momentum and energy conservation in individual scattering processes; he concludes that, if $\gamma$-rays are used to measure the position of a particle, the 
Compton Effect should produce a momentum transfer to the particle of order $\Delta p = h/\lambda$.  The product of electron position uncertainty $\Delta q \sim \lambda$ and the momentum uncertainty produced by the measuring $\gamma$-rays should be $\Delta q \cdot \Delta p \sim h$. If $\gamma$-rays are replaced by other particles "B" as measuring devices, Heisenberg refers to de Broglie and associates a wavelength $\lambda = h/p$ with  particles "B". 
The scattering process again should transfer momentum to particle "A", inducing an uncertainty yielding the same estimate  $\Delta q \cdot \Delta p \sim h$. Higher precision of position requires shorter wavelengths of $\gamma$-rays or higher momenta of particles "B", necessarily  inducing larger momentum uncertainties of particle "A".\\ 

Let us first retain what is valid about  uncertainties and commutation relations: \\
a)  Quantum uncertainties and commutation relations are connected and are consequences of the same physical principle.\\
b) Canonically conjugated variables of a physical system cannot both have arbitrary precision; the product of their relative uncertainties is restricted by a lower bound of order Planck's constant.\\

Heisenberg's "illustrative explanation", however,  is fundamentally flawed; his 
argument, "Observations necessarily cause changes in the system to be measured; these "disturbances" are responsible for quantum uncertainties",
 is incorrect: {\bf Measurements without disturbances in the system to be measured are not only possible, but are carried out 
routinely}\footnote{As has been discussed in the previous chapter, Bragg scattering and all similar phenomena, which had been interpreted to result from "interference effects", require that the scattering system remains unchanged; M\"{o}ssbauer spectroscopy is another example.}. 
Examples for disturbance free measurements of particle positions and their quantum uncertainties will be given in the appendix.\\

Heisenberg confirms that his own understanding is in contradiction to Born-Jordan; Heisenberg writes: Quantum uncertainties may {\it "according to Born and Jordan be viewed as characteristic and statistical elements of quantum theory in contrast to classical theory"}. \\ 
Heisenberg specifies his deviating attitude: {\it "The difference between classical and quantum theory rather consists in: Classically we may assume the phases of the atom to be determined by preceding experiments. In reality, however, this is impossible, because every experiment to determine the phase will destroy or change the phase of the atom."} And concerning "causality": 
{\it  "The sharp formulation of the causality law, "if we know the present exactly, we are able to calculate the future", is not wrong due to the second part of the sentence, but because the precondition is wrong".}\\

This is in sharp contrast to Born's basic principle that all changes in nature are discontinuous and purely statistical. Quantization of the action variable is the key: Discontinuous transitions require the action variable to change by integer values of $h$. Physical systems therefore can no longer be characterized by precise values of physical variables, only averages over corresponding action intervals can be given. 
The basic quantum laws, the commutation relations, are consequences of quantization of the action variable  (Born and Jordan, Z. Phys. 34, 858, 1925); and the uncertainty relations of canonically connected variables are necessary consequences of the commutation relations. The formal derivation from commutation relations to uncertainty relations is contained in the paper by {\bf E. H. Kennard  (Z.Phys. 44, 326, 1927; rec 17 July 1927)} and is given most clearly by {\bf H. P. Robertson (Phys. Rev. 34, 163, 1929)}. \\

No physical variable of any physical system may be perfectly sharp.  Just as the position-momentum commutation relation does not allow a particle to be in an exact momentum eigenstate, the time-energy commutation relation forbids an atom or any other physical system to be in an exact energy eigenstate.
Mathematical models of closed systems may have exact eigenvalues and eigenstates; but these models of closed systems, at best, describe approximations. Physical systems are necessarily open systems, nature always provides additional couplings to "the rest of the world". Taking the relation between energy and time as an example: The additional couplings guarantee that every state of a physical system has finite lifetime, generating a natural linewidth and implying that the energy cannot be perfectly sharp. The time-energy commutation relation imposes a lower bound on the product of energy and lifetime uncertainties.\footnote{Pauli's "energy-representation" (energy real variable and time becomes derivative with respect to energy) proposed in January 1926 is perfectly admissible. The appendix will contain an example. Pauli's "correction" in "Die allgemeinen Prinzipien der Wellenmechanik", Springer Verlag 1933, ("time always has to be treated as ordinary number") is in error. }\\

After submission of Heisenberg's paper, intensive discussions with Bohr continued, which induced Heisenberg to write an addendum.  Bohr had developed ideas, which only partly agreed with Heisenberg's concept, but also contained essential differences. Although Bohr accepted Heisenberg's erroneous argument concerning experiments necessarily provoking disturbances, Bohr advocated a "complementarity principle", particles should have "dual properties", i.e.  particle {\bf and} wave properties. Experimental results observing particles should be due to the "particle property", experiments observing "interference" should be associated with the "wave property" {\bf (N. Bohr, {\it "Das Quantenpostulat und die neuere Entwickling der Atomistk",} Naturwissenschaften 15, 245-257, 1928)}. \\
At first Heisenberg remained 
skeptical\footnote{letter to Pauli of 4 April: {\it "I am quarreling with Bohr whether the relation $\Delta q \Delta p \sim h$ has its origin in the wave- or discontinuity aspect of quantum mechanics. Bohr stresses, that diffraction of waves is essential, I emphasize that the light-quantum theory and the Geiger-Bothe experiment are essential."}};
Schr\"{o}dinger's "waves" should only be calculational tools without physical 
significance\footnote{Concerning "waves" Heisenberg agreed with Born and Jordan, the important disagreement lies in the origin for discontinuities: Heisenberg saw them as the result of disturbances produced by scattering processes, whereas for Born and Jordan they are constitutive elements of quantum physics.}.
The letters to Pauli of 16 May and 25 May 1927 indicate that the controversy led to severe tensions, which were defused somewhat by the addendum; Heisenberg acknowledges Bohr to have pointed out that the wave aspect, in particular the collimation of the $\gamma$-ray microscope, should play an important role for the uncertainty of position measurements.\\

\begin{section}
{General Conclusions and Outlook}
\end{section}

When Max Planck gave the start to the "quantum age", he did so with great hesitation; the introduction of the new fundamental constant, the quantum of action, was made out of mathematical necessity to reproduce the functional form of the "Normalspektrum", replacing the arbitrary assumption of his previous phenomenological derivation. But Planck's mindset concerning natural phenomena was firmly rooted in classical thinking; he hoped that quantization could eventually be replaced by some classical explanation, requiring only minor adjustments about the interaction of radiation with matter. \\

Einstein's proposal to decompose electromagnetic radiation into quantized particles absorbed and emitted as undividable entities constituted a revolution, which - for  Planck and the vast majority of the scientific community -  went too far. Wave behaviour of radiation seemed to be  firmly established and  irreconcilable with Einstein's light quanta. The skepticism towards quantized radiation extended even beyond 1925, when the final breakthrough was achieved by Born, Heisenberg, and Jordan. The application of the new quantum laws to quantization of the electromagnetic field by Born and Jordan (BJ-2) at first was {\it "either totally ignored or viewed as a slight attack of craziness" ("leichter Anflug von Verr\"{u}ckheit"}, Born, “My Life” 1968). When Schr\"{o}dinger proposed “Wave Mechanics” in 1926, not only Schr\"{o}dinger himself, but also Planck and many others  belonging to the old guard, hoped that “Wave Mechanics” could provide a way back to classical physics, eliminating quantization altogether. 
Planck finally conceded that:
{\it "A new scientific truth usually does not gain general acceptance, because its opponents finally declare themselves to be convinced; it is rather that the opponents gradually die out and the new generation is acquainted with the new truth from the start"}. (Max Planck, wissenschaftliche Selbstbiographie, 1948)\\

The "old quantum theory" started from Planck's point of view, retaining classical pictures about mechanics and electrodynamics. The mechanical behaviour of electrons alone was explored; the coupling to the radiation field was left to a postulate, Bohr's "frequency condition".  Bohr tried to guess the intra-atomic dynamics, and, quite naturally, directed his attention towards the simplest possible system, the hydrogen atom. The electron circling the nucleus was a natural choice and Bohr's heuristic quantum conditions were able to reproduce the spectroscopic results. What seemed to be a success at first, again quite naturally, was taken as encouragement to pursue this path further, and more and more assumptions were added to reproduce the spectroscopic data of other elements. Although this path did not lead to the desired breakthrough, the twelve years between 1913 and 1925 were probably necessary to prepare the scientific community for the radical changes ahead. The classical concepts of continuity and determinism, which formed the basis for classical equations of motion, had failed, showing the necessity to look for more radical ideas. \\

It was the failure of the "old quantum theory" which induced  Born to abandon the space-time continuum and to set up the radical "program", laid out in his book "Vorlesungen \"{u}ber Atommechanik" in 1924:  Rejecting all speculations about unobservable intra-atomic dynamics, the "true quantum laws" should contain relations between observable quantities; classical laws should apply to macroscopic averages only; quantum dynamics was to be described by discontinuous and statistical laws; quantum states were to be characterized by quantum numbers; precise values of physical variables should no longer be associated with the elementary constituents of the quantum world! \\

This "program" was conceived before the mathematical formulation of the theory, it was inspired by what had been directly observable about the quantum world. Almost all information available resulted from radiative transitions. The detailed analysis of experimental results about the interaction of radiation with matter had led Einstein to postulate that radiation itself is quantized, consisting of elementary objects having particle character. Einstein had confirmed this hypothesis showing that thermal equilibrium between radiation and matter required the existence of light quanta characterized by their energies and momenta. Finally, Bose showed that Planck's radiation law was fully consistent with the particle concept of light quanta. When Born made the first step towards the "true quantum laws", Einstein's physical concepts defined the direction to follow. But whereas Einstein had refused to take the final step, Born accepted discontinuities and purely statistical behaviour as fundamental principles.\\

The new quantum theory developed by Born, Heisenberg and Jordan relied on quantization of the action variable; the entire theory was constructed from this principle. Quantization of the action variable required discontinuous dynamics; discontinuities implied statistical behaviour. The new quantum laws  contained in commutation relations and quantum theoretical equations of motion were built from these principles. Born introduced {\it "quantum mechanics"}; Born and Jordan recognized that quantization had to apply to all of physics, {\it "quantum mechanics"} and {\it "quantum optics"};
 they attributed the {\it "basic laws of quantum optics"} to Einstein. The new quantum laws were applicable to all physical processes, {\it "quantum mechanics"} and {\it "quantum optics"} and to their mutual coupling. \\

The initial mathematical formulation of the new quantum theory was derived from the preceding physical understanding: Nature is discontinuous and statistical; as was the resulting theory, "Matrix Mechanics". Continuous representations followed quickly, generating great flexibility in mathematical methods, which could be adapted to the specific problems to be treated. Unfortunately, this was not only used as mathematical advantage, it also led to misunderstandings about the physical content.  
Schr\"{o}dinger's reaction, \\
{\it "I felt deterred, if not to say repelled, by the apparently very difficult methods of transcendental algebra and the lack of illustrative clarity",} \\
reflects the attitude of many contemporaries towards the new mode of thought. Discontinuous and statistical behaviour was totally opposite to what Schr\"{o}dinger called {\it "Das r\"{a}umlich zeitliche Denken"} (the mode of thought based on continuity in space and time), and Schr\"{o}dinger concluded: \\
{\it "We are not really able to change the modes of thought and if we cannot understand within these modes of thought, then we cannot understand at all."} \\
Schr\"{o}dinger's point of view has impact until today. Influenced by Schr\"{o}dinger's waves, Bohr proposed "complementarity": Electrons sometimes should behave as particles and sometimes behave as waves;  Bohr himself described this as "irrational". Instead of explaining any basic principle, "complementarity" covered quantum physics with a veil of unnecessary mysticism. \\

When progress in experimental techniques provided access to the quantum world, this was new territory and the old concepts failed to describe the new findings.  Born accepted that old prejudices had to be abandoned; the understanding of the quantum world does require to change the traditional mode of thought in regard to natural phenomena. And that is what Born did, when he abandoned the space-time continuum. The quantum laws derived by Born and Jordan have stood the test of time; the experimental evidence collected over the past 100 years have confirmed their basic principles: Nature on the elementary level is discontinuous and statistical; classical laws are approximately valid for macroscopic averages. \\[20mm]

\noindent
{\large \bf Acknowledgment:} Many thanks to Efim Kats for helpful comments.\\[20mm]

\begin{center}
{\Large \bf Appendix:\\[3mm] Scattering Processes and the Basic Quantum Laws }
\end{center}

The material covered in this appendix is not part of the "elementary" period; the experimental techniques mentioned became available only much later. They are included because scattering experiments are particularly instructive to demonstrate the basic principles of quantum physics; the experimental results can be directly related to the elementary quantum laws, i. e. commutation relations and quantum equations of motion. The measurement of quantum uncertainties of position mentioned in {\bf Section 8} might have surprised the reader; if so, this appendix contains necessary specifications. More details about  Pauli's "energy-momentum representation" might clear up remaining questions about "interference" and the role of "time" in quantum physics. \\

\noindent
The  Hamiltonian
\begin{equation}
H= H_0({\bf \hat p}) + H_0(X) +V( X,{\bf \hat r})
\end{equation}
describes the coupled system of a particle with some other system $X$.  $ H_0({\bf \hat p}) $ is the Hamiltonian of the free particle, $H_0(X)$ the Hamiltonian for the system X. The coupling is taken to be a scalar field $V( X,{\bf \hat r})$. ${\bf \hat p}$ and ${\bf \hat r}$ are particle momentum and position operators with commutation relation
\begin{equation}
[{\bf \hat p \hat r - \hat r \hat p]} = \frac {\hbar}{i} {\bf 1},
\end{equation}
as introduced by Born and Wiener; e. g.  ${\bf \hat p}$ is an operator with eigenvalues ${\bf p}$ and eigenvectors $|{\bf p}\rangle $; explicitly: ${\bf \hat p |p\rangle  = p|p\rangle }$.\\

We  address the experimental setup, where a source provides particles of initial momentum ${\bf p}_i$ to be scattered off a target system described by initial quantum number $X_i$. The scattering process produces final states of momentum ${\bf p}_f$, and target quantum number $X_f$. The momenta are the measured quantities.
Take initial and final states as eigenstates of $H_0({\bf \hat p}) + H_0(X)$. The "equation of motion" for ${\bf \hat p}$ 
\begin{equation}
\langle {\bf p}_f, X_f| \hat{{\bf \dot p}}|X_i,{\bf p}_i\rangle   =  \frac{i}{\hbar}\langle {\bf p}_f, X_f| [H {\bf {\bf \hat p}} - {\bf \hat p} H]|X_i,{\bf  p}_i\rangle 
\end{equation}
reduces to
\begin{equation}
\langle {\bf p}_f, X_f|\hat{{\bf \dot p}}|X,{\bf p}\rangle  = ({\bf p}_i - {\bf p}_f)  \frac{i}{\hbar}\langle {\bf p}_f, X_f| V(X,{\bf \hat r})|X_i,{\bf p}_i\rangle \;.
\end{equation}

Compare the classical and quantum versions of the equation of motion: In classical physics $\dot{{\bf p}}$ is defined by an infinitesimally small momentum interval $d {\bf p}$ divided by an infinitesimally small time interval $d t$ . Quantization of the action variable requires all physical variables to change discontinuously; the quantum equation of motion specifies the admissible transitions. For the example chosen here, the allowed momentum intervals  $\Delta {\bf p} = ({\bf p}_f - {\bf p}_i)$ are determined by the  factor $i/\hbar\langle {\bf p}_f, X_f| V(X,{\bf \hat r})|X_i,{\bf p}_i\rangle $, which has dimension $(dt)^{-1}$; the corresponding transition probability per unit time is proportional to $|\langle {\bf p}_f, X_f| V(X,{\bf \hat r})|X_i,{\bf p}_i\rangle |^2$. \\
If the interaction $V$ is weak enough, the physical scattering process consists of a single elementary quantum transition and is directly described by the quantum equation of 
motion.\footnote{ This is typically the case in neutron scattering; the neutron interacts weakly, whereas particles coupling to electric charges, like photons or electrons,  interact strongly and will often cause multiple quantum transitions. 
In the following we shall refer to scattering processes resulting from a single quantum transition; the qualitative conclusions, however, will also be valid for the scattering of photons and other particles.} \\

\noindent
Initial and final states are product states (e.g. $| X_i,{\bf p}_i\rangle \;=| X_i\rangle |{\bf p}_i\rangle $); we define 
\begin{equation}
V_{X_f,X_i}({\bf \hat r})  = \; \langle  X_f|V(X,{\bf \hat r})|X_i\rangle .
\end{equation}
\noindent
The evaluation of the remaining matrix element $\langle {\bf p}_f |V_{X_f,X_i}({\bf \hat r}) |{\bf p}_i \rangle $ proceeds by Fourier expansion
\begin{equation}
 \langle {\bf p}_f |V_{X_f,X_i}({\bf \hat r}) |{\bf p}_i \rangle  =  \langle {\bf p}_f | \int_{\bf q} \tilde V_{X_f,X_i}({\bf q}) \;e^{- i{\bf  q\cdot \hat r}} |{\bf p}_i \rangle .
\end{equation}
 In momentum representation (${\bf \hat r} = -\frac {\hbar}{i} \vec \nabla_{\bf p}$) the Taylor expansion of $ |{\bf p} + \hbar {\bf q}\rangle $ may be written as $ e^{- i{\bf  q\cdot \hat r}}|{\bf p}\rangle $.
We obtain
\begin{equation}
\langle {\bf p}_i + \hbar {\bf q}, X_f|\dot{{\bf p}}|X_i,{\bf p}_i\rangle  \;= \; i \; {\bf q}\;\tilde V_{X_f,X_i}({\bf q}).
\end{equation}
The transition probability for the particle to be scattered with momentum transfer of  $ \hbar {\bf q}$ and the scattering system making a transition from $X_i$ to $X_f$  is proportional to ${\bf q}^2|\tilde V_{X_f,X_i}({\bf q})|^2$. Although this result may be obtained using any representation, the momentum representation is best suited to make the connection between physical content and mathematical formalism clearest. Initial and final particle momenta are the observable quantities, their representation by real variables avoids the misconception, that wave functions might be more than mathematical tools.\\ 

{\large \bf Bragg scattering}\\

Specify the target system to be a crystal. The total scattering probablity $W_{tot}(X_i)$ for given initial target quantum number $X_i$ is proportional to the sum over all possible final states $|X_f\rangle $.
Of special importance are scattering processes, which do not induce any change in crystal quantum numbers; initial and final state of the crystal are identical, $(X_f = X_i)$, 
\begin{equation}
W_{tot}(X_i) \;\sim \; \int_{\bf q}\;  {\bf q}^2 \bigg(|\tilde V_{X_i,X_i}({\bf q})|^2 + \sum_{X_f \neq X_i} |\tilde V_{X_f,X_i}({\bf q})|^2\bigg).
\end{equation}
The essential condition for Bragg scattering contributions: Only those events, which "do not leave any trace in the crystal" (i. e. the contributions from $X_f = X_i$) may contribute to Bragg scattering, whereas all other scattering (i. e. the sum over $X_f \neq X_i$) contributes to a rather structureless 
background.\footnote{An example for background contributions in neutron scattering: The interaction between neutron spin and nuclear spins contains an additional vectorial coupling, which provides finite probabilities for nuclear spin flips. Transitions at different lattice sites correspond to different final states; accordingly the total scattering probability for nuclear spin transitions is a sum over the scattering probabilities from individual lattice sites. The same argument applies to all processes, which cause localized transitions in the crystal.}\\

Purely elastic scattering events require energy conservation; ${\bf p}_i^2 = {\bf p}_f^2 = ({\bf p}_i + \hbar {\bf q})^2$. Assume that the crystal is in a state of perfect crystalline 
periodicity, characterized by a set of real space vectors ${\bf L}$, such that for integer $n$ 
\begin{equation}
V_{X_i,X_i}({\bf r}+ n{\bf L}) = V_{X_i,X_i}({\bf r}).
\end{equation}
Nonvanishing Fourier coefficients $\tilde V_{X_i,X_i}({\bf q})$ will be restricted to ${\bf q = Q}$, where  

\begin{equation}
{\bf Q \cdot L} = n\; 2\pi, 
\end{equation}
n integer. Discrete translational symmetry selects special momentum transfers $\hbar {\bf Q}$, which, combined with energy conservation  ${\bf p}_i^2 = {\bf p}_f^2 = ({\bf p}_i + \hbar {\bf Q})^2$, characterize the 
Bragg peaks.\footnote{Real samples contain crystalline disorder, which leads to finite widths of Bragg peaks and additional background contributions.} 
Whereas full translational symmetry would require momentum to be conserved, discrete translational symmetry conserves {\bf "quasi momentum"}; a particle of  momentum ${\bf p}$ may be scattered into ${\bf p} + \hbar {\bf Q}$; the scattering probabilities are determined by $|\tilde V_{X_i,X_i}({\bf Q})|^2$.\\

"Quasi momentum conservation" is the direct consequence of "Born's quantization condition":  The products of the symmetry vectors ${\bf L}$ and the allowed momentum transfers $\hbar {\bf Q}$ are equal to the change in action variables, which, according to "Born's quantization condition", have to be  integer multiples of Planck's constant; $\hbar {\bf Q \cdot L} = nh$.\\

{\large \bf Measuring the quantum mechanical position uncertainty}\\

The essential condition for all scattering processes contributing to diffraction peaks is purely elastic scattering, no change in quantum numbers of the crystal may 
occur.
The very existence of diffraction phenomena already disproves  the claim made by Heisenberg and Bohr, that scattering processes necessary cause disturbances in the system to be measured.  
Purely elastic (i. e. {\it "disturbance free"}) scattering processes may actually be used to {\bf measure} the quantum uncertainties of positon. We present a simple example, which has become standard practice in neutron scattering, not only measuring the positions of crystal nuclei but also their quantum uncertainties. \\

The nuclear  interaction between the neutron and the crystal nuclei may  be considered point like and written as a sum over lattice sites $l$
\begin{equation}
V({\bf \hat R}_l,{\bf \hat r}) = \sum_{l}  b_l\; \delta ({\bf \hat r -  \hat R}_l).
\end{equation}
The ${\bf \hat r}$ is the neutron positon operator, the ${\bf \hat R}_l$ are position operators of crystal nuclei, the $b_l$ are the scalar coupling constants. Let the crystal be in a state corresponding to quantum numbers $X_i$ and average positions of nuclei ${\bf R}^i_l$. The diagonal matrix element of the interaction $V({\bf \hat R}_l,{\bf \hat r})$  over $|X_i\rangle $ will result in
\begin{equation}
V_{X_i,X_i}({\bf \hat r})  = \; \sum_{l}  b_l\; f_i ({\bf \hat r -  R}^i_l),
\end{equation}
where the function $ f_i ({\bf  r -  R}^i_l)$ represents the position uncertainty of the nucleus at lattice site $l$ in the state $|X_i\rangle $. The further evaluation may proceed as in the preceding section on Bragg scattering. The purely elastic scattering probability for momentum transfer $\hbar {\bf q}$ is 
\begin{equation}
|\tilde V_{X_i,X_i}({\bf q})|^2 = | \sum_{l}e^{i{\bf q} \cdot {\bf R}^i_l}b_l \cdot\tilde f_{i,l}({\bf q})|^2,
\end{equation}
where the function $\tilde f_{i,l}({\bf q})$  is the Fourier transform of  $f_{i,l}({\bf r})$,  and $|f_{i,l}({\bf r})|^2$ gives the probability distribution for the nuclear position at site $l$ in the state $|X_i\rangle $.  For temperature $T$ tending towards zero  and the crystal being in its ground state ($i = 0$), $f_{0,l}({\bf r})$  represents the nuclear position uncertainty due to zero point fluctuations.\\

For perfect crystalline periodicity the functions $\tilde f_{0,l}({\bf q})$ are identical for equivalent lattice sites; the sum over $l$ on the right hand side of the equation above will guarantee that  
$\tilde V_{X_0,X_0}({\bf q})$ vanishes except for the special values ${\bf q} = {\bf Q}$ (the Bragg peaks). These $\delta$-peaks will attain finite widths in real crystals due to finite grain size of crystallites and crystalline disorder; furthermore, the accuracy of neutron momenta is restricted by experimental resolution and quantum uncertainties. To lowest order, the total intensity of the various Bragg peaks will be unaffected. Their intensities provide a finite number of Fourier components of the functions $f_{0,l}({\bf r})$, and a large enough number of Bragg peaks measured enables a reasonable reconstruction of $f_{0,l}({\bf r})$.\\

At $T = 0$ the position uncertainty is due to zero point quantum uncertainties. For oscillators at $T = 0$ the minimum value of the mean square deviations allowed by the fundamental laws is reached; for finite temperature, states of higher energy are excited, the experiment measures the thermal average and the position uncertainty increases. \\

{\large \bf  Remarks concerning field quantization}\\

Pauli's  "momentum-energy representation" may be used to obtain a {\bf "shortcut" to field quantization}. 
Take the classical interaction between a particle and the field to be of simple scalar form
$V({\bf r}, t)$, where ${\bf r}$ is the position variable of the particle. Although electric and magnetic fields are vector fields and the coupling is not a simple scalar coupling, this is not important for the following; the arguments  below may be applied to quantization of all types of classical fields. \\

Take the Fourier expansion of the classical field $V({\bf r}, t)$:\\

\begin{equation}
 V({\bf r}, t) = \int_{{\bf q},\omega}  \tilde V({\bf q}, \omega) e^{- i (\bf q   r - \omega t)}.
\end{equation}
The transition to quantum theory replaces the classical variables by operators; using Pauli's "momentum-energy representation", ${\bf r}$ is replaced by the operator ${\bf \hat r} = - \hbar /i \;\bf {\nabla_p}$ and $t$ is replaced by $\hat t =  \hbar /i  \cdot  d /d E $. The "perturbation"  $V({\bf \hat r}, \hat t)$ acting on a particle state $|E,{\bf p}\rangle $ of well defined energy $E$ and momentum ${\bf p}$ results in

\begin{equation}
 V({\bf \hat r}, \hat t) \; |E,{\bf p}\rangle  = \int_{{\bf q},\omega} \tilde V({\bf q}, \omega)  \:\; |E + \hbar \omega ,{\bf p} + \hbar {\bf q}\rangle .
\end{equation}
A field of frequency $\nu$ and wavelength $\lambda$ may cause the particle to make a quantum transition with energy transfer $\Delta E = \hbar \omega = h \nu$ and momentum transfer $|\Delta {\bf  p}| = \hbar |{\bf q}| = h/\lambda$; the transition probability  is proportional 
to $|\tilde V({\bf q}, \omega)|^2$. $\Delta E / \nu$ and $\lambda \cdot |{\bf \Delta p}|$ represent the change in action variable, which - as required - is equal to $h$. \\

Quantization of action has to affect {\bf all} physical variables, i. e. must have consequences on the field as well: Born's quantization condition applied to a field of frequency $\nu$ and wavelength $\lambda$ requires the existence of quanta with energy $\epsilon = h\nu$ and momentum $p = h/\lambda$, respecting energy and momentum conservation for the individual elementary quantum transition. This  reflects Einstein's reasoning, when he postulated photons in 1905: The experimental observations of the interaction between radiation and matter, in particular the energy and momentum exchange between the field and point like particles, require the existence of radiation quanta having particle character.  \\

\end{document}